\documentclass[twocolumn]{aastex631}

\usepackage{graphicx}
\usepackage{float}
\DeclareGraphicsExtensions{.pdf}
\usepackage{amsmath,amsfonts,amsthm} % Math packages for equations
\usepackage{aas_macros}
\usepackage[T1]{fontenc}
\usepackage[utf8]{inputenc} % Required for inputting international characters
\usepackage{bm}
\usepackage[english]{babel} % English language hyphenation
\usepackage{acronym}
\usepackage{xspace}
\usepackage{footmisc}

\newcommand{\dd}{\ensuremath{\mathrm{d}}}
\newcommand{\HI}{\text{H}\textsc{i}\xspace}

\newcommand{\Hmol}{\text{H$_2$}\xspace}
\newcommand{\NHI}{\ensuremath{N_{\HI}}}
\newcommand{\XCO}{\ensuremath{X_{\text{CO}}}}
\newcommand{\TB}{\ensuremath{T_{\text{B}}}}
\newcommand{\Ts}{\ensuremath{T_{\text{s}}}}
\newcommand{\TCMB}{\ensuremath{T_{\text{CMB}}}}
\newcommand{\kl}[1]{\left( #1 \right)}

\begin{document}

\title{Diffuse Emission of Galactic High-Energy Neutrinos from a Global Fit of Cosmic Rays}

\author[0000-0002-2050-8413]{Georg Schwefer}
\affiliation{Institute for Theoretical Particle Physics and Cosmology (TTK), RWTH Aachen University, 52056 Aachen, Germany}
\affiliation{III. Physikalisches Institut B, RWTH Aachen University, 52056 Aachen, Germany}
\affiliation{Max-Planck-Institut für Kernphysik, Saupfercheckweg 1, 69117 Heidelberg, Germany}
\correspondingauthor{Georg Schwefer} 
\email{georg.schwefer@mpi-hd.mpg.de}
\author[0000-0002-2197-3421]{Philipp Mertsch}
\affiliation{Institute for Theoretical Particle Physics and Cosmology (TTK), RWTH Aachen University, 52056 Aachen, Germany}
\author[0000-0002-6418-3008]{Christopher Wiebusch}

\affiliation{III. Physikalisches Institut B, RWTH Aachen University, 52056 Aachen, Germany}

\date{\today}

\begin{abstract}
In the standard picture of galactic cosmic rays, a diffuse flux of high-energy gamma-rays and neutrinos is produced from inelastic collisions of cosmic ray nuclei with the interstellar gas.
The neutrino flux is a guaranteed signal for high-energy neutrino observatories such as IceCube, but has not been found yet.
Experimental searches for this flux constitute an important test of the standard picture of galactic cosmic rays. 
Both the observation and non-observation would allow important implications for the physics of cosmic ray acceleration and transport.
We present \texttt{CRINGE}, a new model of galactic diffuse high-energy gamma-rays and neutrinos, fitted to recent cosmic ray data from AMS-02, DAMPE, IceTop as well as KASCADE. 
We quantify the uncertainties for the predicted emission from the cosmic ray model, but also from the choice of source distribution, gas maps and cross-sections.
We consider the possibility of a contribution from unresolved sources.
Our model predictions exhibit significant deviations from older models. 
Our fiducial model is available at \href{https://doi.org/10.5281/zenodo.7373010}{this https URL}.
\end{abstract}

%\maketitle

% ----------------------------------------------------------------------------------------
% ----------------------------------------------------------------------------------------
% ----------------------------------------------------------------------------------------
\section{Introduction\label{sec:introduction}}

Galactic diffuse emission (GDE) of photons and neutrinos is radiation produced within the interstellar medium (ISM) of the Galaxy.
Electromagnetic radiation has been observed at all wavelengths ranging from radio and microwaves up to PeV gamma-rays. 
Of particular interest are photons and neutrinos from hadronic interactions of galactic cosmic rays (GCRs) with the interstellar gas while propagating through the Galaxy \citep{Strong:1998fr}.
Both, diffuse galactic emission of high-energy photons and neutrinos, offer invaluable information on the spatial and spectral distribution of GCRs elsewhere in the Galaxy \citep{Tibaldo:2021viq}, providing immediate information on the century-old problem of the origin of galactic cosmic rays. 

At gamma-ray energies, three processes contribute to the high-energy GDE of photons: decay of neutral pions, produced in inelastic collisions of GCR nuclei with interstellar gas; bremsstrahlung from GCR electrons and positron on the interstellar gas; and Inverse Compton scattering, that is the upscattering of soft radiation backgrounds by GCR electrons and positrons. 
Furthermore, a number of extended structures have been discovered in high-energy gamma-rays, like the Fermi bubbles~\citep{Dobler:2009xz,Su:2010qj,Fermi-LAT:2014sfa} or galactic radio loops~\cite{1971A&A....14..252B}, as well as an isotropic, extragalactic gamma-ray flux \citep{Fermi-LAT:2014ryh}. 
Finally, a certain fraction of the observed diffuse emission is likely due to so-called unresolved sources, that is sources with fluxes below the experimental detection thresholds \citep{Vecchiotti2021a,Vecchiotti2021b}. 
Untangling the various contributions to GDE in gamma-rays is a formidable task that is in part made difficult by the degeneracy between the hadronic and leptonic contributions. 

Unlike the multi-component fluxes of photons,
the flux of galactic diffuse neutrinos offers a clean window to the study of GCR because neutrinos uniquely originate from the decay of charged mesons, i.e.\ pions, that are produced in hadronic interactions. 
This flux is a guaranteed signal for the IceCube Neutrino Observatory \citep{Aartsen2011} and other high-energy neutrino telescopes \citep{Ageron2011,KM3Net:2016zxf,Baikal-GVD:2018isr}, but a discovery has thus far remained elusive. However, recent analyses \citep{IceCube:2019lzm,ANTARES:2018nyb,IceCube:2017trr} suggest that such a discovery could be in reach within the near future.

Precise models of diffuse neutrino emission play a two-fold role in the searches: 
First, they  quantify our expectations for  searches, allowing us to estimate what bearings (non-)observations have for our understanding of models of GCRs. 
Second, they provide detailed spatio-spectral templates  for targeted experimental searches which otherwise suffer from atmospheric and extragalactic backgrounds. 

Model predictions of diffuse galactic neutrino emission can be based on observed photon fluxes because of the close connection between the parent particles, neutral and charged pions, that originate from the same interactions of GCRs.  
Of the models available in the literature, the Fermi-$\pi^0$ model ~\citep{Fermi-LAT:2012edv} and the KRA$\gamma$ model~\citep{Gaggero:2014xla} have been employed in most of the recent experimental searches~\citep{IceCube:2017trr,ANTARES:2018nyb,IceCube:2019lzm}. 
The Fermi-$\pi^0$ model assumes a factorization into the angular distribution of the $\pi^0$-component as modelled by the \textit{Fermi} collaboration and a spectrum that above a few $\mathrm{GeV}$ is a single power law with the same spectral index of $\gamma\approx2.7$ in all directions. 
While designed for use at GeV energies, for neutrino studies, the model spectra have been extrapolated to higher energies with the same unbroken power law. In previous studies, the spectral index has also been left free to float.
The KRA$\gamma$ model instead exhibits harder gamma-ray spectra in the galactic center direction, leading to a different morphology at the energies of interest for observations of high-energy neutrinos
(see also \citet{Luque:2022buq} for a recent update of the KRA$\gamma$ model). 
In addition, there are also a number of analytical parametrizations of fluxes of high-energy gamma-rays and neutrinos~\citep{Joshi:2013aua,Lipari:2018gzn,Fang:2021ylv}.

These existing models, however, suffer from two main drawbacks: 
First, they have not been systematically fitted to the latest data of local GCR observations. 
Given the various uncertainties in the modelling of GDE, local data on GCR should serve as an important anchor point and one would hope that the models are able to reproduce these local data~\citep[e.g.][]{Marinos:2022tdj}. 
Second,  the models suffer from a lack of quantitative estimates of model uncertainties. 
While the sources of such uncertainties are manifold (fit uncertainties from the GCR model, choice of gas maps and cross-sections), both uses of GDE models (see above) rely heavily on a proper estimate of model uncertainties. 

In this paper, we aim at updating existing models in the light of recent data on local GCR fluxes, and present the \texttt{CRINGE}\footnote{\textbf{C}osmic \textbf{R}ay-fitted \textbf{In}tensities of \textbf{G}alactic \textbf{E}mission} model. 
We also quantify the uncertainties from the GCR model parameters and the various choices for GDE inputs. 

While we hope to present a useful state-of-the-art  model for high-energy diffuse galactic neutrinos, we must caution that we rely on existing, \emph{conventional} models for the sources and transport of GCRs. 
Various anomalies in gamma-ray observations have recently pointed to the need for overhauls of such conventional models~\citep{Fermi-LAT:2016zaq,Yang:2016jda}. For instance, the observation of a hardening of the gamma-ray emission towards the inner Galaxy has not been fully understood. 

The outline of the paper is as follows: 
In Sec.~\ref{sec:method}, we start by describing the various ingredients for predictions of the GDE in high-energy neutrinos and gamma-rays. We lay out the GCR model and describe the global fit to local GCR data. 
Next, we discuss the various choices for other inputs of the diffuse model, that is gas maps, cross-sections and the photon backgrounds. 
We also present a model of unresolved sources and explain our treatment of gamma-ray absorption. 
Our results are presented in Sec.~\ref{sec:results}, first for the global fit to local measurements of GCRs, then for the diffuse fluxes of high-energy gamma-rays and neutrinos. 
An extended discussion in the context of other GDE observables can be found in Sec.~\ref{sec:discussion}. 
We conclude in Sec.~\ref{sec:conclusion}.

% ----------------------------------------------------------------------------------------
% ----------------------------------------------------------------------------------------
% ----------------------------------------------------------------------------------------
\section{Method\label{sec:method}}

The intensity of hadronic gamma-rays and neutrinos from longitude and latitude $(l, b)$ and at energy $E$, $J(l, b, E)$, is given as the line-of-sight integral (e.g. \cite{Strong:1998fr}) of the volume emissivity, generated by the inelastic collisions of hadronic GCRs with the gas in the interstellar medium,
\begin{equation}
\begin{aligned}
J(l, b, E) &= \frac{1}{4 \pi} \sum_{m,n} \int_0^{\infty} \mathrm{d} s \int_E^{\infty} \mathrm{d} E' \, \frac{\mathrm{d} \sigma_{m,n}}{\mathrm{d} E}(E', E) \\
& \times J_{m}(\boldsymbol{r}, E') n_{\text{gas},n}(\boldsymbol{r}) \Big|_{\boldsymbol{r} = \boldsymbol{r}(l, b, s)} \, .
\label{eqn:diffuse_intensity_model}
\end{aligned}
\end{equation}
Here, $J_{m}(\boldsymbol{r}, E) = v / (4 \pi) \psi_m (\boldsymbol{r}, p)$ is the GCR intensity of species $m$, where $\psi$ is the isotropic CR density per unit energy . 
$(\mathrm{d} \sigma_{m,n} / \mathrm{d} E)(E', E)$ is the differential cross-section for production of gamma-rays or neutrinos of energy $E$ from inelastic collisions of GCR species $m$ of energy $E'$ on gas of species $n$. 
Finally, $n_{\text{gas},n}(\boldsymbol{r})$ denotes the 3D distribution of gas, mostly atomic and molecular hydrogen in the Galaxy. 

Similarly, the intensity of gamma-rays originating from Inverse Compton scattering of cosmic ray leptons, $J_{\mathrm{IC}}(l, b, E)$, can be calculated as
\begin{equation}
\begin{aligned}
J_{\mathrm{IC}}(l, b, E) &= \frac{1}{4 \pi} \sum_{e^+,e^-} \int_0^{\infty} \mathrm{d} s \int_0^\infty \mathrm{d}E_{0} \int_0^{\infty} \mathrm{d} E' \,\\ &\frac{\mathrm{d} \sigma_{\mathrm{KN}}}{\mathrm{d} E}(E', E ,E_{0})\\ 
& \times J_{e}(\boldsymbol{r}, E') n_{\text{ISRF}}(\boldsymbol{r},E_{0}) \Big|_{\boldsymbol{r} = \boldsymbol{r}(l, b, s)} \,.
\label{eqn:diffuse_IC_model}
\end{aligned}
\end{equation}
Here, $(\mathrm{d} \sigma_{\mathrm{KN}} / \mathrm{d} E)(E', E, E_0)$ is the Klein-Nishina cross-section and $n_{\text{ISRF}}(\boldsymbol{r},E_{0})$ is the number density of radiation field photons per unit energy~\cite{1970RvMP...42..237B}.
The predicted intensity of high-energy gamma-rays or neutrinos as a function of direction and energy therefore depends on these three inputs: a model for the distribution of GCRs, a map of the interstellar gas and radiation field in the Milky Way, and the hadronic production cross-sections. 
In the following, we will detail our modelling choices for each of these inputs. 
We will also describe the global fit that we employed to determine the parameters of our GCR model.

% ----------------------------------------------------------------------------------------
% ----------------------------------------------------------------------------------------
\subsection{Galactic Cosmic Ray Model\label{sec:GCR_model}}

The propagation of GCRs is usually modelled with the transport equation \citep{1965P&SS...13....9P,ginzburg_syrovatskii_1964},
\begin{equation}
\frac{\partial \psi}{\partial t} - \frac{\partial}{\partial x_i} \kappa_{ij} \frac{\partial \psi}{\partial x_j} + u_i \frac{\partial \psi}{\partial x_i} + \frac{\partial}{\partial p} \left( b \frac{\partial \psi}{\partial p} \right) = q \, . 
\label{eqn:TPE}
\end{equation}
The boundary conditions employed are that $\psi$ vanishes on the surface of a cylinder of half-height $z_{\text{max}}$ and radius $r_{\text{max}}$. 
We have assumed $r_{\text{max}} = 20 \, \text{kpc}$ ad adopted $z_{\text{max}}= 6 \, \text{kpc}$~\citep{Evoli:2019iih}. 
The solution of this partial differential equation depends on the assumed transport parameters, that is the spatial diffusion tensor $\kappa_{ij}$, the advection velocity $u_i$, the momentum loss rate $b$ and the source density $q$. 
We  solve eq.~\eqref{eqn:TPE}, employing a publicly available version of the \texttt{DRAGON} code \citep{Evoli2016}, assuming axisymmetry with respect to the direction perpendicular to the galactic disk. 
In the following, we describe the various transport parameters.

% ----------------------------------------------------------------------------------------
\subsubsection{Diffusion\label{sec:diffusion}}

GCRs diffuse due to resonant interactions with a spectrum of turbulent magnetic fields~\cite{Berezinsky1990}. 
Here, we assume isotropy, such that the diffusion tensor is a diffusion coefficient $D$ times unit matrix, $\kappa_{ij} = D \delta_{ij}$. 
If the power spectrum of turbulence was known, the diffusion coefficient $D$ could be computed. 
In phenomenological applications, however, the diffusion coefficient is oftentimes assumed to follow a certain parametric form. 

As for its rigidity-dependence, we employ a power law with four breaks, 
\begin{equation}
\begin{aligned}
D(\mathcal{R})&=D_0\beta\kl{\frac{\mathcal{R}}{\mathcal{R}_{12}}}^{-\delta_{1}}\\& \times\prod_{i=1}^{4} \kl{1+\kl{\frac{\mathcal{R}}{\mathcal{R}_{i(i+1)}}}^{1/s_{i(i+1)}}}^{-s_{i(i+1)}(\delta_{i+1}-\delta_i)}
\label{eq:diffkoeff_model}
\end{aligned}
\end{equation}
hence four break rigidities $\{\mathcal{R}_{i(i+1)}\}$, four softness parameters $\{s_{i(i+1)}\}$, five spectral indices $\{\delta_i\}$ and one normalization $D_0$. 

Of course, such spectral breaks should only be introduced if required for an accurate description of the data 
(see ~\citet{Vittino2019}, where the necessity of spectral breaks in the diffusion coefficient is discussed for models of GCR electrons and positrons).
At the same time, it is important to motivate the breaks from a physical mechanism, e.g.\ from features in the power spectrum of turbulence. 
In the following, we provide some pointers as to the physical origin of the breaks we consider. 
While the exact break parameters will be determined from a global fit to local GCR data, we indicate some benchmarks.

At rigidities of a few GV and above, the propagated spectra are to a first approximation proportional to the source spectrum divided by the diffusion coefficient. 
Under the constraint of producing the observed spectral indices, the spectral indices of the source spectra and of the diffusion coefficient are therefore approximately degenerate. 
This degeneracy gives us the freedom to choose a pure power law for the source spectra, see below, and instead absorb possible spectral breaks in the source spectrum into breaks of the diffusion coefficient. 

The first spectral break ($\delta_2-\delta_1 > 0$) in the diffusion coefficient at a few GV is a hardening of its spectrum and serves to absorb a spectral softening of the source spectrum. 
Such spectral breaks have been observed in the gamma-ray spectra of supernova remnants (SNRs) by \textit{Fermi}-LAT~\citep{Fermi-LAT:2009qzy,FermiLAT:2009kcy,Fermi-LAT:2010kue,Fermi-LAT:2010rdy}. 
One possibility is that the break in the diffusion coefficient is due to self-confinement of GCRs in the near-source environment~\citep{Jacobs:2021qvh}. 
Such a break has also been introduced into models of GCRs on purely phenomenological grounds, that is in order to achieve a better fit to locally measured spectra when not including reacceleration, as is the case for the model constructed here. 
We note that even for a model that includes reacceleration such a break is likely necessary~\citep{Strong:2011wd}. 

The hardening break in the GCR spectra ($\delta_3-\delta_2 < 0$) at $\mathcal{R}_{23} \simeq 300 \, \mathrm{GV}$ has been found to be present both in primary~\citep{Panov:2006kf,Ahn:2010gv,PAMELA:2011mvy,AMS:2015tnn,AMS:2015azc,AMS:2017seo} and secondary GCRs~\citep{AMS:2018tbl}. 
The fact that the break is more pronounced in secondary species points to a propagation effect~\citep{Vladimirov:2011rn,Genolini:2017dfb} rather than a feature in the source spectrum. 
It has been shown~\citep{Blasi:2012yr,Evoli:2018nmb} that such a break can be rather naturally explained as a transition in the turbulence power spectrum from self-generated turbulence dominating $D$ for low rigidities to external turbulence for high rigidities.

A further softening in the GCR spectra ($\delta_4-\delta_3 > 0$) has been observed in the spectra of proton and helium by the DAMPE~\citep{DAMPE:2019gys,Alemanno:2021gpb} and CALET~\citep{CALET:2019bmh,Brogi:2021csp} experiments. 
Earlier indications from the CREAM experiment exist~\citep{Yoon:2017qjx}. 
While it has been suggested to be the spectral feature of an individual nearby source~\citep{Malkov:2020smm,Fornieri:2020kch}, statistically such a scenario is considered  unlikely~\citep{Genolini:2016hte}. 
Instead, it might be attributed to a cut-off of one population of sources, e.g.\ supernova remnants (SNRs), before a different population takes over at higher rigidities. 

Finally, the softening ($\delta_5-\delta_4 > 0$) break in the GCR spectra around $\sim \text{PV}$ is the well-known cosmic ray ``knee''. 
Although it has been discovered in the all-particle spectrum of cosmic rays  as early as 1959~\citep{Kulikov_Khristiansen_1959}, there is still no consensus about its origin.
The KASCADE and KASCADE-Grande experiments have identified it to be consistent with a break at fixed rigidity for different elements \citep{KASCADE:2005ynk,KASCADEGrande:2011kpw}.
Therefore the leading hypotheses are that it corresponds either to the maximum rigidities at which galactic magnetic fields can contain GCRs  or to the maximum rigidity of galactic accelerators of cosmic rays~\citep{1984ARA&A..22..425H}.

% ----------------------------------------------------------------------------------------
\subsubsection{Source Injection\label{sec:source_injection}}

We follow the usual assumption of a factorization of the source term $q=q(\boldsymbol{r},E)$ in eq.~\eqref{eqn:TPE} into a spatial source distribution $S(r,z)$ and an injection spectrum, $g(p)$, 
\begin{equation}
\label{eq:injection_split}
Q(\boldsymbol{r},p) = S(r,z)g(p) \, .
\end{equation}

The spatial source distribution $S(r,z)$ is an input to the cosmic ray model with a significant impact on the morphology of the resulting diffuse emission. 

To estimate the associated uncertainty, we use the four commonly used distributions from \citet{Ferriere2001}, \citet{Case1998}, \citet{Lorimer2006} and \citet{Yusifov2004}. All of these are analytical parametrizations based on population studies of SNRs, massive stars as progenitors or pulsars as relicts of supernovae in the Milky Way. Thus, they all serve as a proxy for the distribution of SNRs, the likely preeminent sources of galactic cosmic rays.
For an early inference of the radial source distribution from diffuse GeV gamma-rays, see~\citet{1977ApJ...217..843S}. 

While overall all four parametrizations agree qualitatively, there is significant quantitative disagreement between them. This is true in particular towards the galactic center, where the source density of the \citet{Case1998} and \citet{Yusifov2004} models is forced to zero, while the distributions of \citet{Ferriere2001} and \citet{Lorimer2006} attain finite values.

\begin{figure}

\includegraphics[scale=1]{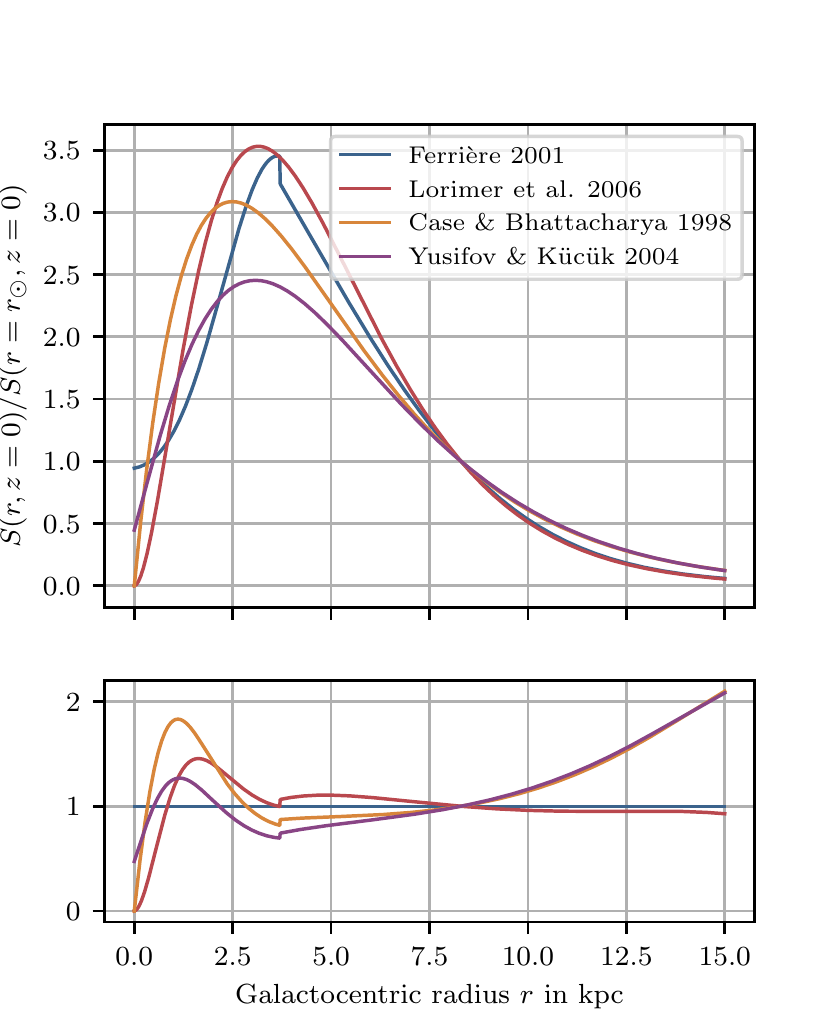}
\caption{
Radial profiles and relative differences to the \citet{Ferriere2001} model of the four GCR source distributions used in this work.
}
\label{fig:sd_radial_profiles}
\end{figure}

The injection spectra $g_i(p)$ for all nuclei $i$ are assumed to be pure power laws with energy-independent, but in general different, spectral indices, that is $g(p) \propto p^{-\gamma_i}$. 
Possible breaks in the source spectra can to a certain extent be absorbed into breaks in the diffusion coefficient, see the discussion in Sec.~\ref{sec:diffusion}. 

For the calculation of the diffuse emission, the contributions of cosmic ray nuclei heavier than helium can be approximated via a scaling factor in the hadronic production cross-sections for $p$-$p$ interactions as in \citet{Casandjian2015}, significantly reducing the computation time. This is because the relevant quantity for the production of hadronic diffuse emission is the all-nucleon flux as a function of kinetic energy per nucleon, $E_{\text{kin}}/n$. As $E_{\text{kin}}/n\propto (Z/A) \mathcal{R}$, spectral features at a common rigidity $\mathcal{R}$ appear also at similar $E_{\text{kin}}/n$ for all nuclei. Therefore, the spectra of nuclei heavier than helium feature a similar $E_{\text{kin}}/n$-dependence to those of lighter nuclei and their contribution to the all-nucleon flux remains small at all $E_{\text{kin}}/n$. 

Because of this, the all-nucleon flux in our model is dominated by the contributions from protons and helium at all $E_{\text{kin}}/n$. 
It also allows us to approximate the subdominant contributions of cosmic ray nuclei heavier than helium via a scaling factor in the hadronic production cross-sections for $p$-$p$ interactions as in \citet{Casandjian2015}, significantly reducing the computation time needed for the calculation of the diffuse emission.

Besides the three spectral indices $\gamma_{\text{p}}$, $\gamma_{\text{He}}$ and $\gamma_{\text{C}}$, the source abundances of helium $N_{\text{He}}$ and carbon $N_{\text{C}}$ relative to those of protons, which, as is for example done in the supplementary material\footnote{\href{https://galprop.stanford.edu/PaperIISuppMaterial/}{https://galprop.stanford.edu/PaperIISuppMaterial/}} to \cite{Fermi-LAT:2012edv}, is set arbitrarily to $1.06\times10^6$, are also free parameters in the model.

For the lepton spectra, the situation is somewhat more complicated. 
Following~\citet{Vittino2019}, we model the injection spectrum of electrons as a twice broken power law, with the spectral indices $\gamma^{e^{-}}_{i}$ and break energies $E^{e^{-}}_{i(i+1)}$ as free fit parameters.
To account for the very short propagation distances of cosmic ray electrons at TeV energies, which are not correctly grasped by the assumed smooth source distribution, the injection spectrum is exponentially cut off at $E^{e^{-}}_{\text{cut}}=20 \, \mathrm{TeV}$. 
This is similar to the choices made in~\citet{Mertsch2018}, where such a cut-off was found to be necessary to match the TeV $e^{-}+e^{+}$ data. 

To account for the spectral hardening in the positron spectrum at $\mathrm{GeV}$ energies~\citep{AMS:2019rhg}, an extra source component yielding equal amounts of electrons and positrons is added to the model. 
Similarly to~\citet{Vittino2019}, it represents additional astrophysical lepton sources, but is agnostic of any precise models for these sources. 
The assumed spatial distribution is the same as for all other sources. 
The spectrum of this extra component is modelled as a once broken power law with an exponential cut-off. 
The break energy is fixed to $E^{\text{extra}}_{12}=50 \, \mathrm{GeV}$, the cut-off is at $E^{\text{extra}}_{\text{cut}}=600 \, \mathrm{GeV}$. 
The two spectral indices $\gamma^{\text{extra}}_i$ are free fit parameters.

% ----------------------------------------------------------------------------------------
% ----------------------------------------------------------------------------------------
\pagebreak
\subsection{Interstellar Medium Components\label{sec:gas_maps}}

Computing the diffuse emission of neutrinos and hadronic gamma-rays requires a 3D map $n_{\text{gas}}(\boldsymbol{r})$ of the gas in the Galaxy, see eq.~\eqref{eqn:diffuse_intensity_model}. 
Such 3D maps can be obtained from galactic surveys of gas line emission. 
Those are based on the fact that galactic rotation induces different relative velocities between the gas and the observer. 
Assuming a velocity model, e.g.\ a rotation curve, this can be used to convert the survey into a 3D map. 
Given the uncertainties of such a reconstruction, models with rather larger bins in galacto-centric radius were developped, so-called ring models. 
(See appendix~B of~\citet{Fermi-LAT:2012edv} for some details.) 
More recently, more sophisticated Bayesian inference techniques have been used~\citep{Mertsch2021b,Mertsch2022}. 
Alternatively, analytical parametrizations of the spatial gas distributions have been suggested \citep{Lipari:2018gzn} as well as more complicated parametrizations that were fit to data~\citep{Johannesson:2018bit}. 
Our choices for the maps of both atomic and molecular gas are described in Secs.~\ref{sec:atomic_hydrogen} and \ref{sec:molecular_hydrogen}.

For the computation of leptonic Inverse Compton gamma-ray emission, a model of the ISRF of the Milky Way is required. We describe the models used in this work in Sec.~\ref{sec:ISRFs}.

% ----------------------------------------------------------------------------------------
\subsubsection{Atomic Hydrogen}
\label{sec:atomic_hydrogen}

Atomic hydrogen (\HI) is traced by the well-known $21 \, \mathrm{cm}$ emission line from the hyperfine transition. 
Combining the data from various telescopes, the LAB survey~\citep{Kalberla2005} and the more recent HI4PI survey~\citep{BenBekhti2016} have become available. 
The quantity measured by these surveys is the \textit{brightness temperature} $\TB(l,b,v_{\text{LSR}})$ of the emission line as a function of direction and radial velocity $v_{\text{LSR}}$ with respect to the local standard of rest. 
The transformation into a differential column density $(\dd \NHI / \dd v_{\text{LSR}})(l,b,v_{\text{LSR}})$ can be calculated from the thermodynamics of two-level systems~\citep{Draine2010,Fermi-LAT:2012edv}
\begin{equation}
\label{eq:HI_T_to_column}
\frac{\dd \NHI}{\dd v_{\text{LSR}}} = C \Ts \tau = - C \Ts \ln{\kl{1-\frac{\TB}{\Ts - \TCMB}}}.
\end{equation}
with $C = 1.823\times10^{18} \, \mathrm{cm}^{-2} \, (\mathrm{K} \, \mathrm{km} \, \mathrm{s}^{-1})^{-1}$.
This involves the spin temperature $\Ts$, which is equivalent to the population ratio of the excited state to the ground state of the hyperfine structure transition. 
A large $\Ts$, corresponding to a large population of the excited state and in consequence little self-absorption and a small optical depth, therefore results in less gas column density being inferred from the observed $\TB$ of the emission. 
In fact, the limit $\Ts \rightarrow \infty$ corresponds to the optically thin limit and poses a lower limit on the total amount of $\HI$~\citep{Mertsch2022}. 
Correspondingly, low $\Ts$ lead to higher inferred column densities.
$\Ts$ is quite uncertain and also thought to vary across the Galaxy~\citep{Fermi-LAT:2012edv,BenBekhti2016}. 
Typically, models assume a constant value of either $\Ts \rightarrow \infty$ or $\Ts \approx 100-500 \, \mathrm{K}$, corresponding to a regime where optical depth becomes relevant (see for example \cite{Fermi-LAT:2012edv,Mertsch2022}).

In order to estimate the uncertainty associated with the choice of the gas distributions, we employ three different models to calculate diffuse emission. These are

\begin{itemize}
\item \texttt{GALPROP}: This is the gas map described as $^{\mathrm{T}}150^{\mathrm{C}}5$ in \cite{Fermi-LAT:2012edv}. It is based on the the LAB \HI survey data deconvolved into 17 rings and assumes $\Ts =150\;\mathrm{K}$. It also includes a correction for dark gas calculated from dust reddening maps \citep{Fermi-LAT:2012edv,Grenier2005}. 
\item \texttt{GALPROP}-OT: This is exactly the same distribution as the \texttt{GALPROP} model, except that the gas is assumed to be optically thin at a spin temperature of $\Ts =10^5\;\mathrm{K}$. In \citet{Fermi-LAT:2012edv}, this model is called $^{\mathrm{T}}100000^{\mathrm{C}}5$.
\item \texttt{HERMES}: This model uses the HI4PI \HI survey data deconvolved into 11 rings and assuming $T_{Spin}=300\;\mathrm{K}$ \citep{Remy2022}.\footnote{See also \href{https://fermi.gsfc.nasa.gov/ssc/data/analysis/software/aux/4fgl/Galactic_Diffuse_Emission_Model_for_the_4FGL_Catalog_Analysis.pdf}{https://fermi.gsfc.nasa.gov/ssc/data/analysis/software/\linebreak aux/4fgl/Galactic\_Diffuse\_Emission\_Model\_for\_the\_4FGL\linebreak\_Catalog\_Analysis.pdf}}
\end{itemize}

% ----------------------------------------------------------------------------------------
\subsubsection{Molecular Hydrogen}
\label{sec:molecular_hydrogen}

Maps of molecular hydrogen (\Hmol) have been derived from the CfA survey compilation~\citep{Dame2000} of emission of the $J=1 \to 0$ line of carbonmonoxide~\citep{Heyer2015}. 
Similar to the $21 \, \mathrm{cm}$ surveys, it provides a map of brightness temperatures $\TB(l,b,v_{\text{LSR}})$. 
The conversion to the differential column density $(\dd N_{\Hmol{}} / \dd v_{\text{LSR}})(l,b,v_{\text{LSR}})$ is however not as rigorous as the procedure in eq.~\eqref{eq:HI_T_to_column}. 
Instead, a phenomenological conversion factor $\XCO$ is used to obtain the following relation
\begin{equation}
\frac{\dd N_{\Hmol}}{\dd v_{\text{LSR}}}=\XCO \TB.
\end{equation}
There is great uncertainty associated with the $\XCO$ factor: 
While locally a value around $2\times10^{20} \, \mathrm{cm}^{-2} \, (\mathrm{K} \, \mathrm{km} \, \mathrm{s}^{-1})^{-1}$ is generally recommended~\citep{Bolatto2013}, different works assume values as low as $0.5\times10^{20} \, \mathrm{cm}^{-2} \, (\mathrm{K} \, \mathrm{km} \, \mathrm{s}^{-1})^{-1}$~\citep{Liu2022} and as high as $8\times10^{20} \, \mathrm{cm}^{-2} \, (\mathrm{K} \, \mathrm{km} \, \mathrm{s}^{-1})^{-1}$~\citep{Luque:2022buq}. 
Also, $\XCO$ is often taken to vary with galactocentric radius, with different models assuming different radial dependencies. 
A common description is a ring-based model, where $\XCO$ is constant within different galactocentric rings. 
Examples of this are~\citet{Gaggero:2014xla}, where $\XCO$ is fixed in two rings, and~\citet{Fermi-LAT:2012edv}, where it is left as a free fit parameter in 13 rings. 
In these models, $\XCO$ typically increases with galactocentric radius. 
This is explained through a decrease in metallicity in the outer Galaxy that increases the $\Hmol$ to CO ratio~\citep{Gaggero:2014xla}.

In this study, we conservatively opt for a constant value of $X_{CO}=2\times10^{20}\;\mathrm{cm}^{-2}\;(\mathrm{K}\;\mathrm{km}\;\mathrm{s}^{-1})^{-1}$ throughout the galaxy, in accordance with the above mentioned local recommendations.

\subsubsection{Interstellar Radiation Field\label{sec:ISRFs}}

Models of the interstellar radiation field (ISRF) in the Milky Way, require, besides the well-measured homogeneous cosmic microwave background (CMB), calculations of the accumulated galactic starlight and the infrared emission from dust. These calculations are challenging as the absorption and re-emission of starlight from dust couples these components. Different analytical \citep[e.g.][]{Vernetto2016} and numerical \citep[e.g.][]{Porter2008,Porter:2017vaa} approaches are used, yielding quantitatively different results. To account for this uncertainty in the calculation of Inverse Compton gamma-ray fluxes, we consider two different models, namely those from \citet{Porter2008} (henceforth called \texttt{GALPROP)} and \citet{Vernetto2016}.

% ----------------------------------------------------------------------------------------
% ----------------------------------------------------------------------------------------
\subsection{Hadronic Production Cross-sections\label{sec:cross-sections}} 

The calculation of hadronic gamma-ray and neutrino production cross-sections in different hadronic interaction models is associated with sizeable uncertainties, as was for example recently detailed by \citet{Koldobiskiy2021}. 
To estimate the resulting uncertainties on the diffuse emission models, we have used three different models for our calculations. These are
\begin{itemize}
\item \text{K\&K:} This model combines the parametrization of the total inelastic $p$-$p$ cross-section from~\citet{Kafexhiu2014} with the secondary yields and spectra from~\citet{Kelner2006}. 
The latter is based on an analytical parametrization of the \texttt{SIBYLL} hadronic interaction model \cite{Fletcher:1994bd}. Note that this parametrization is only valid for primary energies above $100 \, \mathrm{GeV}$~\citep{Kelner2006}. 
\item \text{KamaeExtended:} For primary energies below $500 \, \mathrm{TeV}$, this model uses the parametrization from~\citet{Kamae2006} which is in large parts derived from the \texttt{PYTHIA 6.2} event generator. 
At higher energies, it is extended with the K\&K model. This follows the prescription used in~\citet{Gaggero2015}.
\item \texttt{AAfrag:} This is based on interpolation tables from the hadronic interaction model \texttt{QGSJET-II-04m} described in~\citet{Kachelriess2019,Koldobiskiy2021}. 
Below $4 \, \mathrm{GeV}$ primary energy, it is complemented by the parametrization from~\citet{Kamae2006}.
\end{itemize}

The K\&K and KamaeExtended parametrizations are only available for $p$-$p$ interactions~\citep{Kamae2006, Kelner2006} and interactions of heavier gas and cosmic ray nuclei need to be described via scaling factors of the $p$-$p$ cross-sections. 
AAfrag in principle contains explicit models for the interactions of heavier nuclei~\citep{Koldobiskiy2021}. 
In this work, these are however treated through the same scaling factors as used for the other cross-section parametrizations. 
The scheme used for these scaling factors directly follows~\citet{Casandjian2015}: 
For protons and helium cosmic ray nuclei, which are included in the sum in eq.~\eqref{eqn:diffuse_intensity_model}, the scaling factors relative to the $p$-$p$ cross-sections for different targets are taken from~\citet{Mori2009}. 
As already described in Sec.~\ref{sec:source_injection}, interactions of cosmic ray nuclei heavier than helium are not calculated explicitly, but are rather treated through an increase in the rate of $p$-$p$ interactions. 
 
% ----------------------------------------------------------------------------------------
% ----------------------------------------------------------------------------------------
\subsection{Unresolved Sources}

Individual sources of high-energy gamma-rays or neutrinos vary in luminosity (also known as intrinsic brightness) and in distance from the observer. 
The resulting fluxes (also know as apparent brightnesses) can therefore vary over many orders of magnitude. 
Observations are, however, flux-limited due to a number of effects, like the presence of backgrounds or source confusion. 
A source with a flux smaller than a certain threshold value can therefore not be detected with the required significance and all such sources contribute collectively to the observed diffuse flux, even though the emission is originating from spatially well-defined regions and not from the interstellar medium as for the truly diffuse emission produced by GCRs. 
These sources and their collective flux are commonly referred to as ``unresolved sources''. 

We have modelled the flux of unresolved sources by extrapolating spectra and luminosity distributions from $\gamma$-ray observations at TeV energies. 
We largely follow \citet{Vecchiotti2021a} who considered the unresolved sources to be dominated by pulsar-powered sources. 
While it might appear that such sources would be leptonic, we remain agnostic as to the nature of the particles producing the high-energy gamma-rays inside the sources. 
Later, when modelling the flux of high-energy neutrinos, we also consider a contribution from the same unresolved, pulsar-powered sources as in gamma-rays. Here, we briefly summarize the salient points of the model and enumerate the adopted parameter values~\citep{Vecchiotti2021a}. 

The intensity from unresolved sources of high-energy gamma-rays of energy $E$ observed from the direction $(l, b)$ is the cumulative intensity up to the threshold $J_{\text{th}}$ in intensity, 
\begin{equation}
J_{\text{unres}}(E, l, b) = \int_0^{J_{\text{th}}} \dd J \, p_J(J; E, l, b) \, J  \, , \label{eqn:unresolved1}
\end{equation}
where $p_J$ is the probability density for the intensity $J$ at energy $E$ and from the direction $(l, b)$. 
We assume that the source density factorizes into a volume density of sources $S_*(\boldsymbol{r})$ and a probability distribution of source rates $p_{\Gamma}(\Gamma)$. 
Under this condition, eq.~\eqref{eqn:unresolved1} can be shown to lead to 
\begin{equation}
\begin{aligned}
J_{\text{unres}}(E, l, b) &= \varphi(E) \int_{L_{\text{min}}}^{L_{\text{max}}} \dd L \, \frac{\dd N}{\dd L} \frac{\Gamma}{4 \pi} \\ 
& \times \int_{\sqrt{\Gamma/(4 \pi \Phi_{\text{th}})}}^{\infty} \dd s \, S_*(\boldsymbol{r}(s, l, b)) \, . 
\end{aligned}
\end{equation}
Here, $\varphi(E)$ denotes the spectrum of a single source, 
\begin{equation}
\varphi(E) = \frac{\beta - 1}{1 - 100^{1-\beta}} \left( \frac{E}{1 \, \text{TeV}} \right)^{-\beta} \exp \left[ - \frac{E}{E_{\text{cut}}} \right] \, ,
\end{equation}
normalized to $\sim 1$ in the energy range $1$ to $100 \, \text{TeV}$, assuming a cut-off energy $E_{\text{cut}} = 500 \, \text{TeV}$. 
For the luminosity function $\dd N / \dd L$, we follow again \citet{Vecchiotti2021a} in adopting a power law form, 
\begin{equation}
\frac{\dd N}{\dd L} = \frac{\Gamma_* \tau (\alpha - 1)}{L_{\text{max}}} \left( \frac{L}{L_{\text{max}}} \right)^{-\alpha} \, ,
\end{equation}
with $\Gamma_* = 1.9 \times 10^{-2} \, \text{yr}^{-1}$, $\tau = 1.8 \times 10^3 \, \text{yr}$, $\alpha = 1.5$ and $L_{\text{max}} \equiv L_{\text{max,HESS}} = 4.9 \times 10^{35} \, \text{erg} \, \text{s}^{-1}$. 
The source rate $\Gamma$ can be related to source luminosity $L$ as $\Gamma = (\beta -2)/(\beta - 1) L$. 
Throughout, we have assumed a spectral index $\beta = 2.3$. 
We have varied the flux threshold $\Phi_{\text{th}}$ between $0.01$ and $0.1 \, \Phi_{\text{Crab}}$ where for the 1 to $100 \, \text{TeV}$ energy range $\Phi_{\text{Crab}} = 2.26 \times 10^{-11} \text{cm}^{-2} \, \text{s}^{-1}$. 
Finally, for the spatial distribution $\rho_*(\boldsymbol{r})$ we have assumed the distribution from \citet{Lorimer2006} and $\boldsymbol{r}(s, l, b)$ denotes the position at a distance $s$ from the observer in the direction $(l, b)$. 
The per-flavor neutrino intensity $J_\nu(E_\nu)$ is related to this gamma-ray intensity $J_{\gamma}(E_\gamma) = J_{\text{unres}}(E_\gamma)$ as
\citep[e.g.][]{Ahlers2014,Fang:2021ylv}
\begin{equation}
    J_\nu(E_\nu)=2J_\gamma(2E_\nu) \, .
\end{equation}

% ----------------------------------------------------------------------------------------
% ----------------------------------------------------------------------------------------
\subsection{Gamma-Ray Absorption}

Above a few $\mathrm{TeV}$, gamma-rays are subject to absorption in photon-photon interactions with the ISRF during propagation in the Milky Way. In a fully self-consistent treatment, this is accounted for through the inclusion of the survival probability $\mathrm{exp}(-\tau(E,s(l,b))$ in eq.~\eqref{eqn:diffuse_intensity_model}, where, following e.g. \citet{Vernetto2016}, the optical depth $\tau$ is calculated as the line-of-sight integral
\begin{equation}
\tau(E,s(l,b))=\int_{0}^{s(l,b)} \dd s' \, K(\boldsymbol{r}(l, b, s'),E).
\end{equation}
This depends on the absorption coefficient $K({\boldsymbol{r}},E)$, which for an isotropic ISRF with an photon density per unit energy $n_{\mathrm{ISRF}}(\boldsymbol{r},E_{0})$ is calculated as 
\begin{equation}
K({\boldsymbol{r}},E)=\int \dd E_{0}\, n_{\mathrm{ISRF}(\boldsymbol{r},E_{0})}\langle \sigma_{\gamma\gamma}(E,E_{0}) \rangle 
\end{equation}
with the interaction cross-section $\sigma_{\gamma\gamma}$ averaged over the angle between the two interaction partners.

The overall effect of the absorption thus depends on the assumed galactic distributions of cosmic rays and gas, the ISRF model and the direction in the sky. The dependence on both the choice of ISRF model and details of the cosmic ray and gas distributions is however only weak \citep{Vernetto2016,Breuhaus2022}, in parts because the dominant contribution to the absorption stems from interactions with the homogeneous CMB.

Therefore, and because the fully self-consistent calculation as layed out above adds another line-of-sight integration to eq.~\eqref{eqn:diffuse_intensity_model} and is thus computationally expensive, we use a simplified approach instead. In this, we obtain the absorbed gamma-ray intensity averaged over a given window in the sky $\Omega$, $J_{\mathrm{abs}}(E,\Omega)$ from the non-absorbed intensity $J$ as 
\begin{equation}
J_{\mathrm{abs}}(E,\Omega)=p_{\mathrm{abs}}(E,\Omega)J(E,\Omega)   
\end{equation}
with a separately calculated absorption probability $p_{\mathrm{abs}}(E,\Omega)$. We calculate this following the prescriptions described in App. E of \citet{Breuhaus2022}. We also assume the same ISRF model \citep{Popescu2017} and analytical descriptions of the distributions of galactic gas \citep{Ferriere1998,Ferriere2007} and cosmic rays \citep{Lipari:2018gzn} as used there\footnote{We are grateful to to Mischa Breuhaus for providing his implementation of the calculation, which makes use of the GAMERA code \citep{Hahn2016}.}.

% ----------------------------------------------------------------------------------------
% ----------------------------------------------------------------------------------------
\subsection{Global Fit}

The GCR model described in Sec.~\ref{sec:GCR_model} contains a total of 26 free parameters describing the modeling of sources and transport of GCRs, see Table~\ref{tab:freeparam}. 
We determine those by fits to local measurements of GCR intensities, which requires an additional set of 8 parameters that are also listed in Table~\ref{tab:freeparam}.
In the following, we describe these additional parameters, the datasets considered, the numerical tools used as well as the procedure adopted for the global fit. 

% ----------------------------------------------------------------------------------------
\subsubsection{Data}

We consider both data from direct measurements by the space experiments AMS-02 and DAMPE as well as indirect measurements from IceTop and KASCADE. 
Specifically, we fit to protons~\citep{AMS:2015tnn}, helium~\citep{AMS:2017seo}, carbon~\citep{AMS:2017seo}, electrons~\citep{AMS:2019iwo}, positrons~\citep{AMS:2019rhg}, and the boron-to-carbon ratio from AMS-02~\citep{AMS:2018tbl}, and to the proton~\citep{DAMPE:2019gys} and helium~\citep{Alemanno:2021gpb} data from DAMPE. 
For IceTop~\citep{IceCube:2019hmk} and KASCADE~\citep{KASCADE:2005ynk} data, we use the proton and helium analyses based on the SIBYLL-2.1 interaction model. 

The measurements of the hadronic intensities are however not necessarily consistent between different experiments. 
The most commonly cited reason for this are uncertainties in the (relative) energy scale calibrations of different experiments, in particular for indirect observations such as those by IceTop and KASCADE, but also for calorimetric experiments such as DAMPE~\citep{Adriani2022}. 
We here follow \citet{Dembinski:2017zsh} in introducing one nuisance parameter $\alpha_k$ per experiment $k$ in order to rescale the experimental intensities. 
For instance, given a reported intensity $J(E)$ as a function of reported energy $E$, we rescale it according to
\begin{equation}
\tilde{J}(\tilde{E}) = \frac{1}{\alpha_k} J(\alpha_k \tilde{E}) \, .
\end{equation}
The nuisance parameters $\{\alpha_k\}$ are also determined by the fit. We impose a log-normal prior with a width of $30\,\%$ on each $\alpha_k$. For AMS-02 and DAMPE, the data are sufficiently constraining that the choice of prior has no influence (see Table~\ref{tbl1}).

Finally, we treat solar modulation in the force-field model~\citep{Gleeson1968}. 
We allow for four different modulation potentials, for protons, nuclei, electrons and positrons, $\phi_p$, $\phi_{Nuc}$, $\phi_{e^-}$ and $\phi_{e^+}$. 

% ----------------------------------------------------------------------------------------
\pagebreak
\subsubsection{Numerical Tools}

We have used a publicly available version of the \texttt{DRAGON} code~\citep{Evoli:2008dv} that solves the transport eq.~\eqref{eqn:TPE} with a finite difference method. In this, we have implemented our multi-break model for the diffusion coefficient (eq.~\eqref{eq:diffkoeff_model}) and the parameters associated with it (see Table~\ref{tbl1}).

For computing the hadronic diffuse emission, we have employed the \texttt{HERMES} code~\citep{Dundovic2021}. 
This code provides a flexible framework for computing the volume emissivities, given the spatially resolved GCR spectra and gas densities, see eq.~\eqref{eqn:diffuse_intensity_model}. 
In addition to the models and parametrizations provided with the publicly available version of \texttt{HERMES}, we have added the above mentioned \texttt{GALPROP} and \texttt{GALPROP}-OT gas maps, the \texttt{AAfrag} cross-section parametrization and respective interfaces to the code.

% ----------------------------------------------------------------------------------------
\subsubsection{Procedure}

For fitting our model to the observational data, we adopt a Gaussian likelihood function, combining the quoted statistical and systematic uncertainties of each AMS-02 and DAMPE in quadrature. For KASCADE and IceTop, we have used the statistical uncertainty only and have subsumed the additional systematic uncertainties into a potentially larger energy scale uncertainty.
Given the large number of free parameters, finding the best fit is a non-trivial task. 
It has been recognized that conventional optimizers cannot guarantee to find the global minimum of the negative log-likelihood. 
Instead, it has been suggested to use Markov Chain Monte Carlo (MCMC) techniques for the minimization~\citep{Korsmeier:2016kha,Mertsch:2020ldv}. 
While computationally rather expensive, MCMC samplers are much more robust and less prone to getting stuck in local minima. 
At the same time, once the MCMC chain has converged, the ensemble of samples can be used as an estimate of the parameter credible intervals. 

We have employed an affine-invariant method~\citep{2010CAMCS...5...65G} as implemented in the \texttt{emcee} package~\citep{Foreman-Mackey:2012any}. 
The inherent parallel nature allows for a significant speed-up compared to serial MCMC samplers. Overall, 90000 MCMC samples were drawn, with a single \texttt{DRAGON} calculation taking around 12 minutes on a single core. 
In order to speed up the convergence towards the global minimum, we distinguish between parameters for the \texttt{DRAGON} code (``slow parameters'') and other parameters, that is the energy rescalings $\{\alpha_i\}$ and the force-field potentials (``fast parameters''). 
The ``slow'' parameters are sampled by the MCMC method while the ``fast'' parameters are profiled over after each \texttt{DRAGON} run.
For all further calculations, including the estimates of the uncertainties on the GCR observables shown below, we use the final 6000 samples drawn after convergence of the MCMC chain.

% ----------------------------------------------------------------------------------------
% ----------------------------------------------------------------------------------------
% ----------------------------------------------------------------------------------------
\section{Results\label{sec:results}}

% ----------------------------------------------------------------------------------------
% ----------------------------------------------------------------------------------------
\subsection{Galactic Cosmic Ray Intensities}

\begin{figure*}[thb]
\centering
\includegraphics[scale=1,trim={0 22.14cm 20.96cm 0}, clip=true]{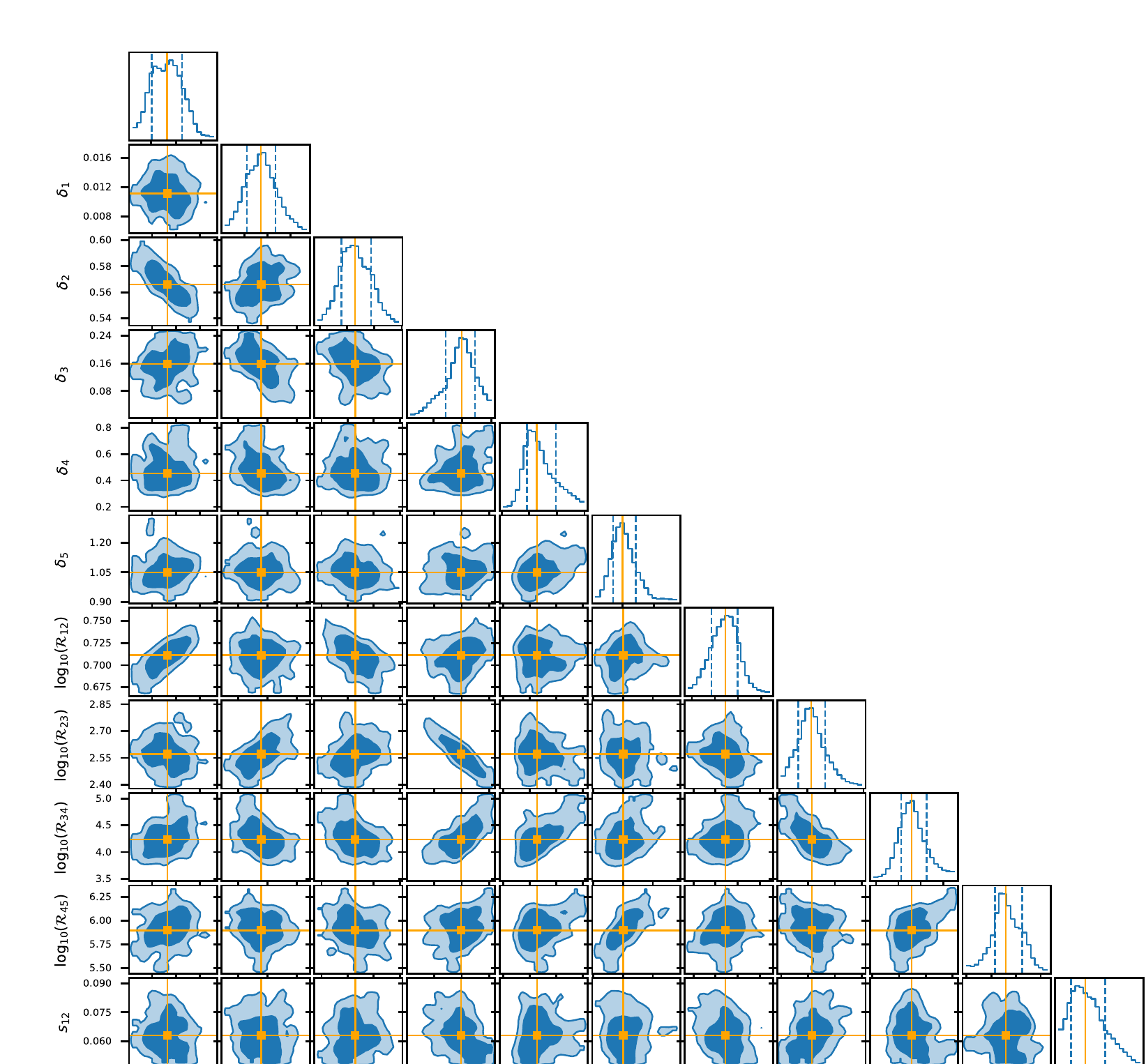} \\
\vspace{-0.04cm}
\includegraphics[scale=1,trim={0 0.8cm 20.96cm 35.76cm}, clip=true]{corner_final_1400_smooth_1_with_quantiles_median_correct_labels.pdf} \\
\caption{Corner plot of the marginalized distributions of the ``slow'' parameters of our model. The values of the break rigidities $\mathcal{R}_{i(i+1)}$ are assumed to be in $\text{GV}$. The dark blue regions are the the estimated $68\,\%$ credible intervals of the 2D marginalized distributions. Similarly, the $68\,\%$ credible intervals in the 1D marginalized distributions are indicated by the dashed blue lines. The light blue regions represent the estimated $95\,\%$ credible intervals of the 2D marginalized distributions. The orange markers indicate the median value for each parameter, which we adopt as our best-fit parameter value. } 
\label{fig:corner1}
\end{figure*}

\begin{figure*}[p]
\centering
\includegraphics[scale=1,trim={0.7cm 0.8cm 20.96cm 15.57cm}, clip=true]{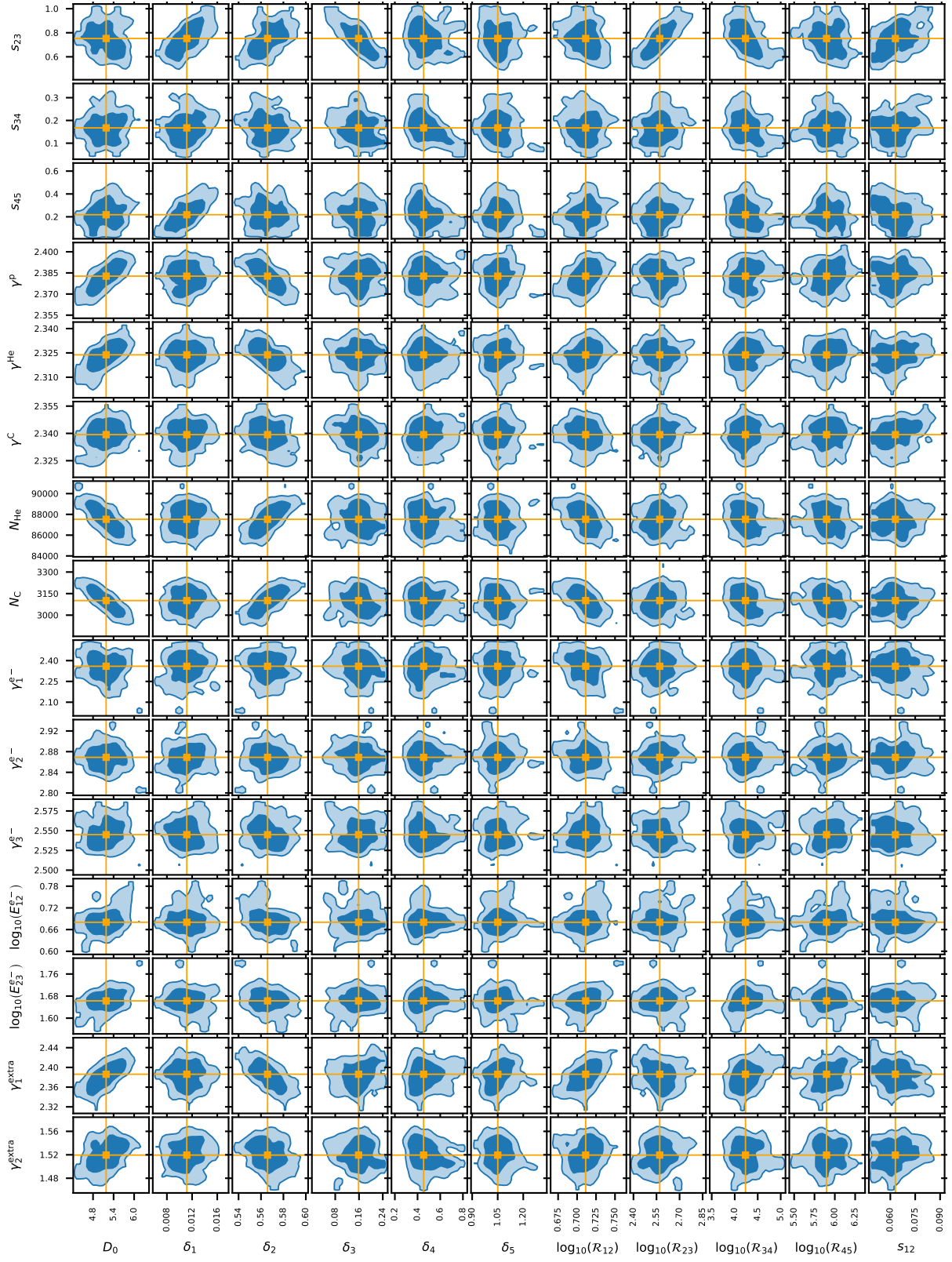}
\caption{Corner plot (cont'd from Figure~\ref{fig:corner1}).} 
\label{fig:corner2}
\end{figure*}

\begin{figure*}[p]
\centering
\includegraphics[scale=1,trim={0.7cm 0.8cm 35.76cm 15.57cm}, clip=true]{corner_final_1400_smooth_1_with_quantiles_median_correct_labels.pdf} \hspace{-0.7em}
\includegraphics[scale=1,trim={16.75cm 0.8cm 6.14cm 15.57cm}, clip=true]{corner_final_1400_smooth_1_with_quantiles_median_correct_labels.pdf}
\caption{Corner plot (cont'd from Figures.~\ref{fig:corner1} and ~\ref{fig:corner2}).}
\label{fig:corner3}
\end{figure*}

\begin{figure}[thb]
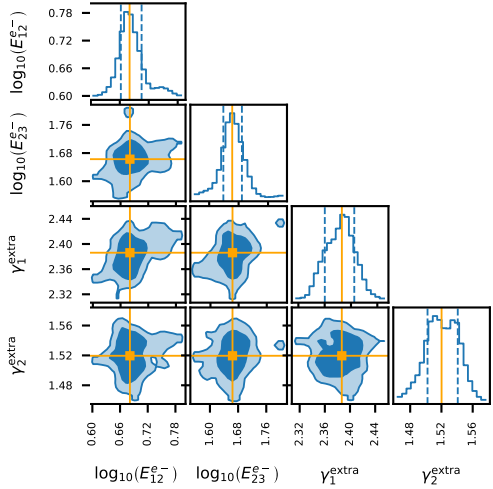

\centering
\includegraphics[scale=1,trim={0.7cm 1.85cm 35.76cm 30.4cm}, clip=true]{corner_final_1400_smooth_1_with_quantiles_median_correct_labels.pdf} \hspace{-0.7em}
\includegraphics[scale=1,trim={31.58cm 1.85cm 0 30.4cm}, clip=true]{corner_final_1400_smooth_1_with_quantiles_median_correct_labels.pdf} \\
\vspace{-0.04cm} \hspace{0.973cm}
\includegraphics[scale=1,trim={31.5cm 0.8cm 0 35.8cm}, clip=true]{corner_final_1400_smooth_1_with_quantiles_median_correct_labels.pdf}
\caption{Corner plot (cont'd from Figs.~\ref{fig:corner1}, \ref{fig:corner2} and \ref{fig:corner3}). The values of the break energies $E^{e-}_{i(i+1)}$ are assumed to be in $\text{GeV}$.}
\label{fig:corner4}
\end{figure}

Figures~\ref{fig:corner1}, \ref{fig:corner2}, \ref{fig:corner3} and \ref{fig:corner4} show (different parts of) the corner plot, that is the collection of the one-dimensional marginal distributions of parameters and of two-dimensional distributions of pairs of parameters.
We have made use of the \texttt{corner} package \citep{corner} adopting a value of $1$ for the \texttt{smooth} and \texttt{smooth1d} keywords.
This is made necessary by the high dimensionality of our parameter space.
Here, we have chosen to show the distributions for the ``slow'' parameters of the MCMC scan only, that is those parameters which are input parameters for the GCR propagation code. 
We remind the reader that for each set of those ``slow'' parameters, we have determined the ``fast'' parameters, that is the remaining ones, through profiling, meaning the likelihood is maximized with respect to the ``fast'' parameters only while the ``slow'' parameters are kept fixed.

For instance, while most of the parameters are uncorrelated, some of the existing correlations and anti-correlations are noteworthy: 
\begin{itemize}
\item The better known anti-correlations in GCR parameters are those between source spectral indices and indices of the diffusion coefficient. 
This is best seen for the spectral indices in the rigidity range where data are most constraining. 
Indeed, the spectral index $\delta_2$ of the diffusion coefficient for $\mathcal{R}$ between $\mathcal{R}_{12}$ and $\mathcal{R}_{23}$ is anti-correlated with the source spectral indices of various species, that is $\gamma^{\text{p}}$, $\gamma^{\text{He}}$ and $\gamma_1^{\text{extra}}$, see Figure~\ref{fig:corner2}.
\item These  anti-correlations between source spectral indices and diffusion coefficient spectral indices also induce correlations between different source spectral indices, see for instance the correlation of $\gamma^{\text{p}}$ and $\gamma^{\text{He}}$ in Figure~\ref{fig:corner3}. 
\item Another apparent correlation is the one between the normalization $D_0$, defined as the value of the diffusion coefficient at the break rigidity $\mathcal{R}_{12}$, and this break rigidity, see Figure~\ref{fig:corner1}. 
\item The spectral indices of the diffusion coefficient above and below a break can  be correlated or anti-correlated with the break rigidity. 
For instance, the break at $\mathcal{R}_{23}$ is a softening of the diffusion coefficient, that is $(\delta_3 - \delta_2) < 0$. 
Increasing $\mathcal{R}_{23}$ can thus be compensated to a certain degree by making the spectral index above, $\delta_3$, even smaller. 
This explains the anti-correlation between $\delta_3$ and $\mathcal{R}_{23}$, seen  in Figure~\ref{fig:corner1}. 
\item The correlation between $\delta_3$ and $\mathcal{R}_{34}$ instead, see  Figure~\ref{fig:corner1}, is due to the fact that the data would prefer a smaller $\mathcal{R}_{34}$ were $\delta_3$ chosen to be smaller. 
\end{itemize}

In Table~\ref{tbl1}, we list the best-fit values of the various ``slow'' and ``fast'' parameters as well as their $68 \, \%$ credible intervals.\footnote{Here, we employ the Bayesian terminology even though strictly speaking, it only applies to unbiased samples from the posterior distribution.} 
We have found the best-fit values to coincide with the maxima and medians of the marginal distributions. 
For the ``slow'' parameters, we have determined the edges of the credible intervals as the $16 \, \%$ and $84 \, \%$ quantiles of the marginal distributions. The edges of the credible intervals for the ``fast'' parameters are similarly calculated as the $16 \, \%$ and $84 \, \%$ quantiles of their distribution over all samples.

\begin{table*}[!th]
\caption{Free parameters and their best-fit values.\label{tab:freeparam}}
\small
\label{tbl1}
\begin{tabular}{r l l}
\hline\hline
\multicolumn{3}{c}{Source parameters} \\
\hline
$\gamma^{\mathrm{p}}$ 			& $= {2.383}^{+0.008}_{-0.009}$ 	& Proton source spectral index \\
$\gamma^{\mathrm{He}}$				& $= {2.324}^{+0.006}_{-0.007}$	& Helium source spectral index \\
$\gamma^{\mathrm{C}}$ 		& $= {2.339}^{+0.006}_{-0.006}$ 	& Carbon source spectral index \\
$N_{\mathrm{He}}$ 						& $= {87520}^{+1170}_{-1200}$ 					& Helium source abundance for a fixed proton abundance of $N_{\mathrm{p}}=1.06 \times 10^6$\\
$N_{\mathrm{C}}$				& $= {3101}^{+69}_{-80}$ 					& Carbon source abundance for a fixed proton abundance of $N_{\mathrm{p}}=1.06 \times 10^6$ \\
$\gamma^{e^-}_1$				& $= {2.359}^{+0.065}_{-0.087}$ 					& Electron source spectral index below $E^{e-}_{12}$ \\
$\log_{10}\left[ E^{e-}_{12} / \text{GeV} \right]$ 	& $= {0.680}^{+0.025}_{-0.018}$ 					& Energy of first break in electron source spectrum\\
$\gamma^{e^-}_2$						& $= {2.869}^{+0.018}_{-0.020}$					& Electron source spectral index between $E^{e-}_{12}$ and $E^{e-}_{23}$ \\
$\log_{10}\left[ E^{e-}_{23} / \text{GeV} \right]$ 		& $= {1.663}^{+0.026}_{-0.024}$ 					& Energy of second break in electron source spectrum \\
$\gamma^{e^-}_3$		& $= {2.545}^{+0.016}_{-0.013}$ 					& Electron source spectral index above $E^{e-}_{23}$ \\
$\gamma^{\mathrm{extra}}_1$ 		& $= {2.386}^{+0.019}_{-0.026}$ 					& Spectral index of extra lepton component below $E^{\text{extra}}_{12}=50 \, \mathrm{GeV}$ \\
$\gamma^{\mathrm{extra}}_2$ 							& $= 1.520^{+0.021}_{-0.018}$ 						& Spectral index of extra lepton component above $E^{\text{extra}}_{12}=50 \, \mathrm{GeV}$ \\
\hline \multicolumn{3}{c}{Transport parameters} \\
\hline
$D_0$ 						& $= {5.18}^{+0.36}_{-0.37}$ 					& Normalization of diffusion coefficient in 10$^{28}$ cm$^{2}$s$^{-1}$ \\
$\delta_1$						& $= {0.0116}^{+0.0019}_{-0.0019}$ 					& Spectral index of diffusion coefficient below $\mathcal{R}_{12}$ \\
$\log_{10}\left[ \mathcal{R}_{12}/ \text{GV} \right]$				& $= {0.711}^{+0.013}_{-0.015}$ 					& First break rigidity \\
$s_{12}$				& $= {0.0630}^{+0.0098}_{-0.0072}$					& Softness of break at $\mathcal{R}_{12}$ \\
$\delta_2$						& $= {0.566}^{+0.012}_{-0.010}$ 					& Spectral index of diffusion coefficient between $\mathcal{R}_{12}$ and $\mathcal{R}_{23}$ \\
$\log_{10}\left[ \mathcal{R}_{23}/ \text{GV} \right]$				& $= {2.571}^{+0.073}_{-0.069}$ 					&Second break rigidity \\
$s_{23}$				& $= {0.75}^{+0.10}_{-0.11}$					& Softness of break at $\mathcal{R}_{23}$ \\
$\delta_3$						& $= {0.159}^{+0.036}_{-0.044}$ 					& Spectral index of diffusion coefficient betwen $\mathcal{R}_{23}$ and $\mathcal{R}_{34}$ \\
$\log_{10}\left[ \mathcal{R}_{34}/ \text{GV} \right]$				& $= {4.23}^{+0.27}_{-0.19}$ 					& Third break rigidity \\
$s_{34}$				& $= {0.167}^{+0.050}_{-0.052}$					& Softness of break at $\mathcal{R}_{34}$ \\
$\delta_4$						& $= {0.453}^{+0.141}_{-0.070}$ 					& Spectral index of diffusion coefficient between $\mathcal{R}_{34}$ and $\mathcal{R}_{45}$ \\
$\log_{10}\left[ \mathcal{R}_{45}/ \text{GV} \right]$				& $= {5.89}^{+0.16}_{-0.11}$ 					&Fourth break rigidity \\
$s_{45}$				& $= {0.022}^{+0.11}_{-0.07}$					& Softness of break at $\mathcal{R}_{45}$ \\
$\delta_5$						& $= {1.050}^{+0.063}_{-0.046}$ 					& Spectral index of diffusion coefficient above $\mathcal{R}_{45}$ \\
\hline \multicolumn{3}{c}{Solar modulation parameters} \\
\hline
$\phi_p$						& $= 0.781^{+0.029}_{-0.036}$ 						& Fisk potential for protons in GV \\	
$\phi_{e^+}$					& $= 0.610^{+0.014}_{-0.017}$ 						& Fisk potential for positrons in GV \\	
$\phi_{e^{-}}$					& $= 1.039^{+0.031}_{-0.037}$ 						& Fisk potential for electrons in GV \\	
$\phi_{\mathrm{nuc}}$			& $= 0.795^{+0.027}_{-0.043}$ 						& Fisk potential for nuclei in GV \\	
\hline \multicolumn{3}{c}{Energy scale shift parameters} \\
\hline
$\alpha_{\mathrm{AMS-02}}$						& $= 1.021^{+0.023}_{-0.022}$ 						& Energy scale shift for AMS-02 \\	
$\alpha_{\mathrm{DAMPE}}$					& $= 1.037^{+0.025}_{-0.022}$ 						& Energy scale shift for DAMPE \\	
$\alpha_{\mathrm{KASCADE}}$					& $= 0.880^{+0.132}_{-0.092}$ 						& Energy scale shift for KASCADE \\	
$\alpha_{\mathrm{IceTop}}$			& $= 0.584^{+0.060}_{-0.084}$ 						& Energy scale shift for IceTop \\	
\hline\hline
\end{tabular}
\end{table*}

\begin{figure*}[p]
\centering
\includegraphics[width=0.49\textwidth]{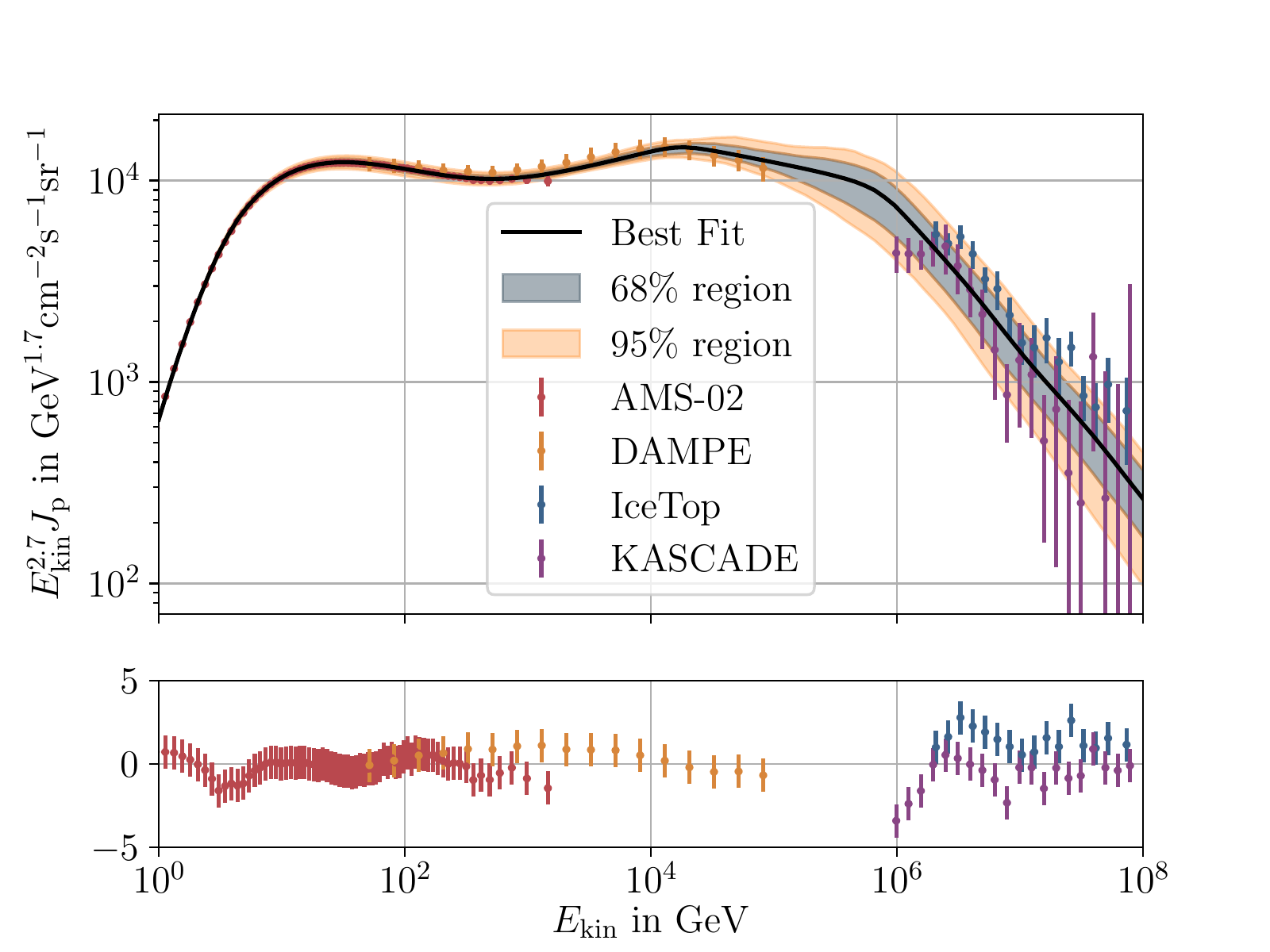} \includegraphics[width=0.49\textwidth]{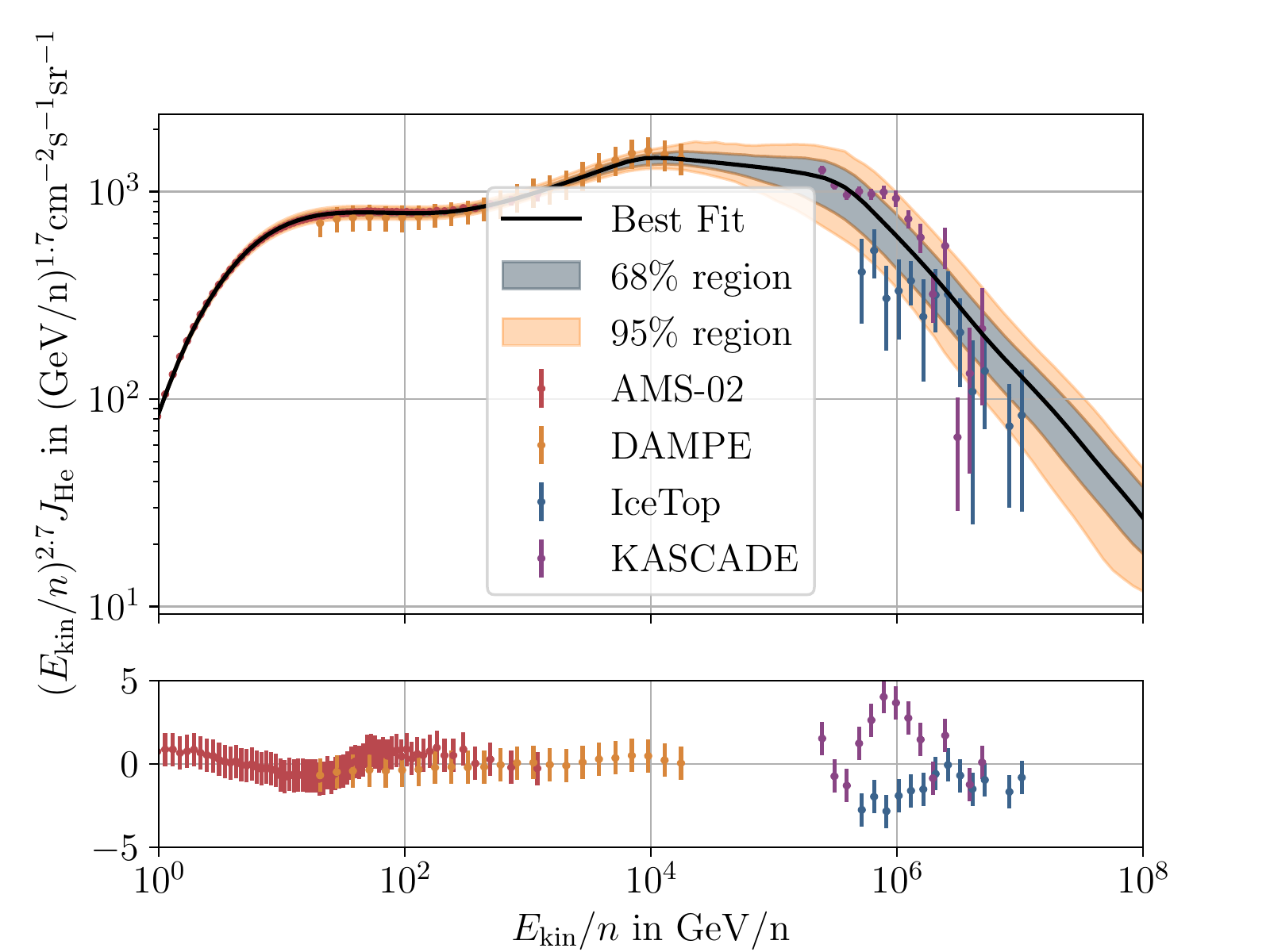} 
\includegraphics[width=0.49\textwidth]{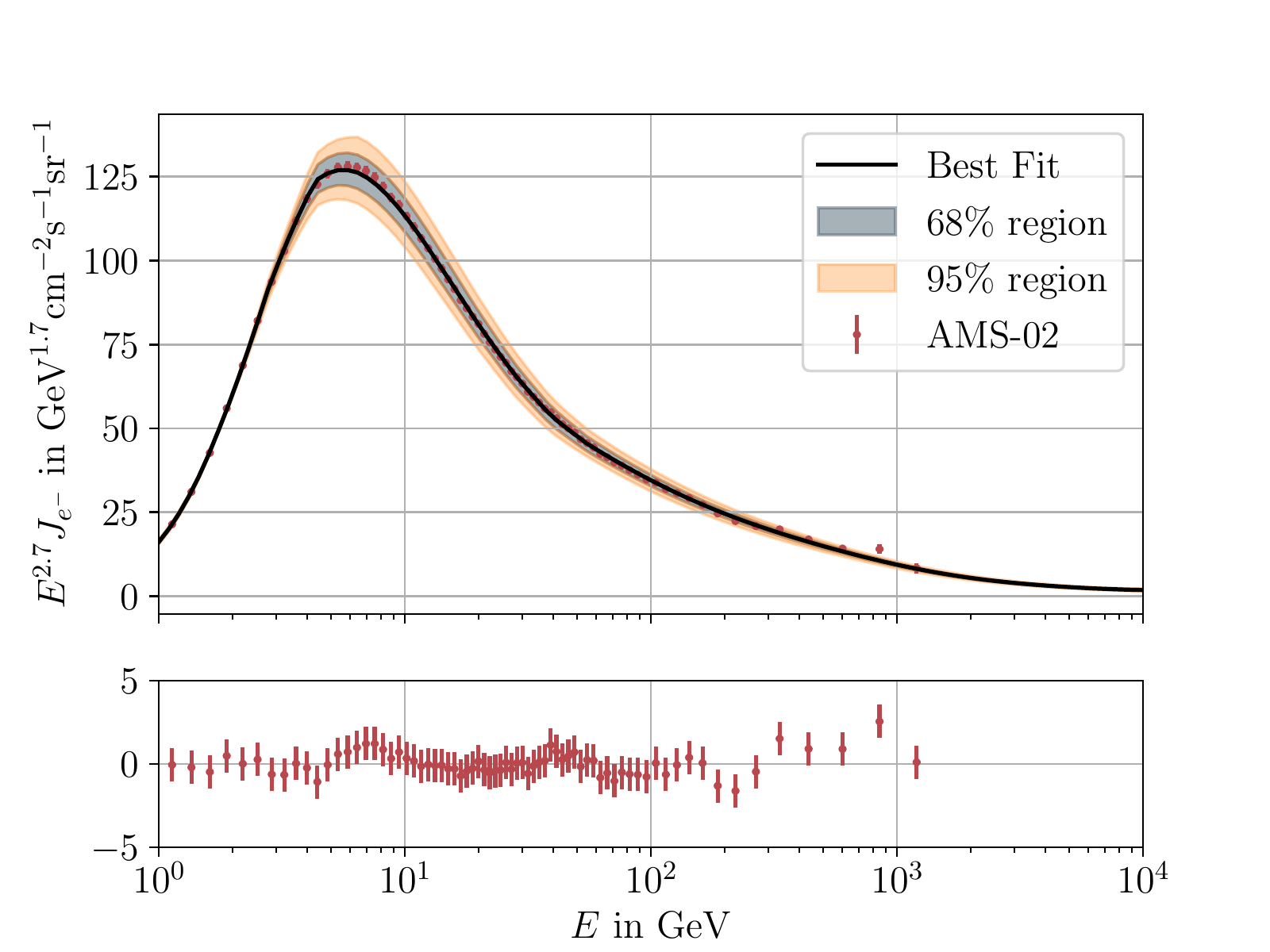} \includegraphics[width=0.49\textwidth]{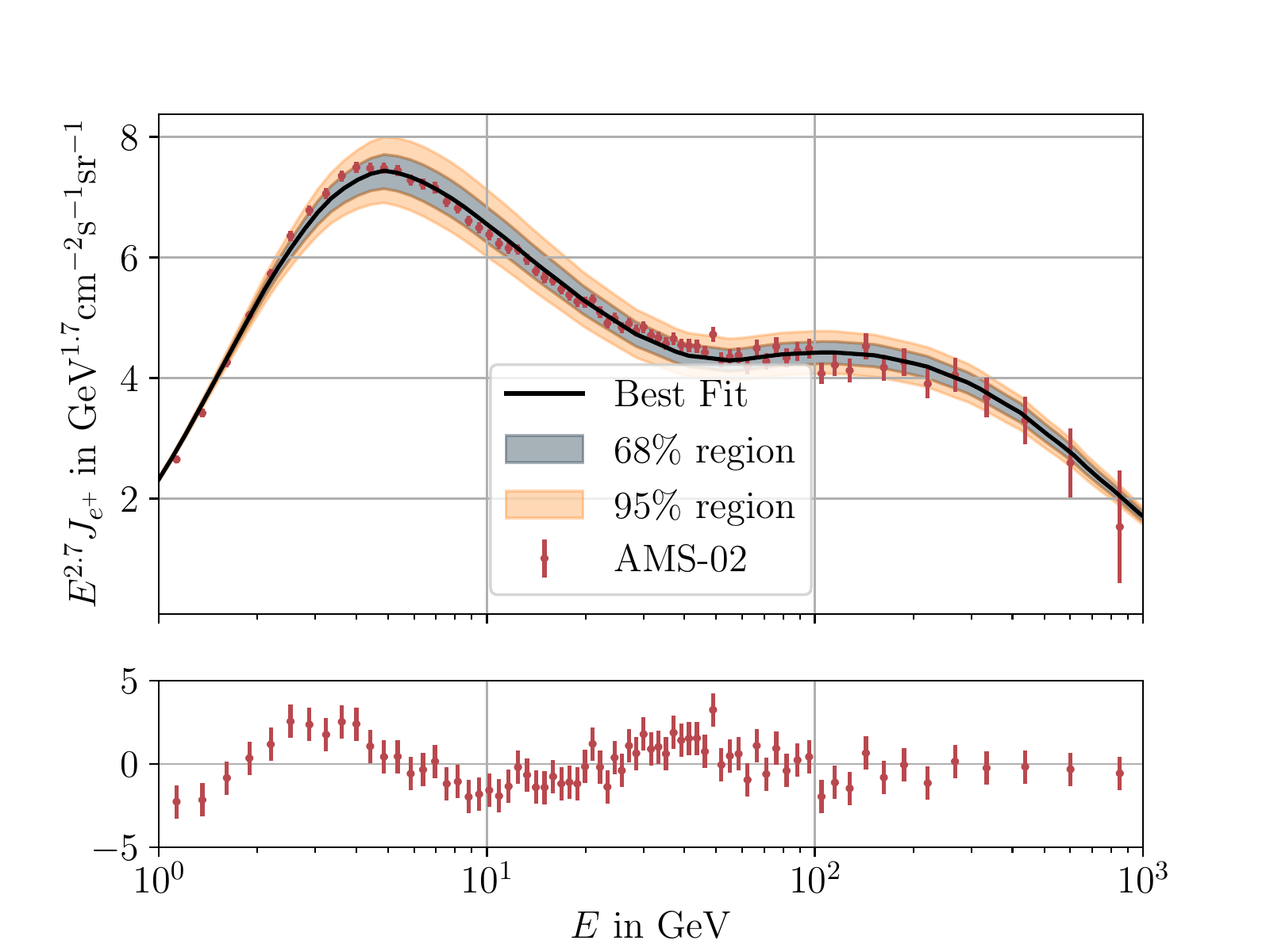} \\
\includegraphics[width=0.49\textwidth]{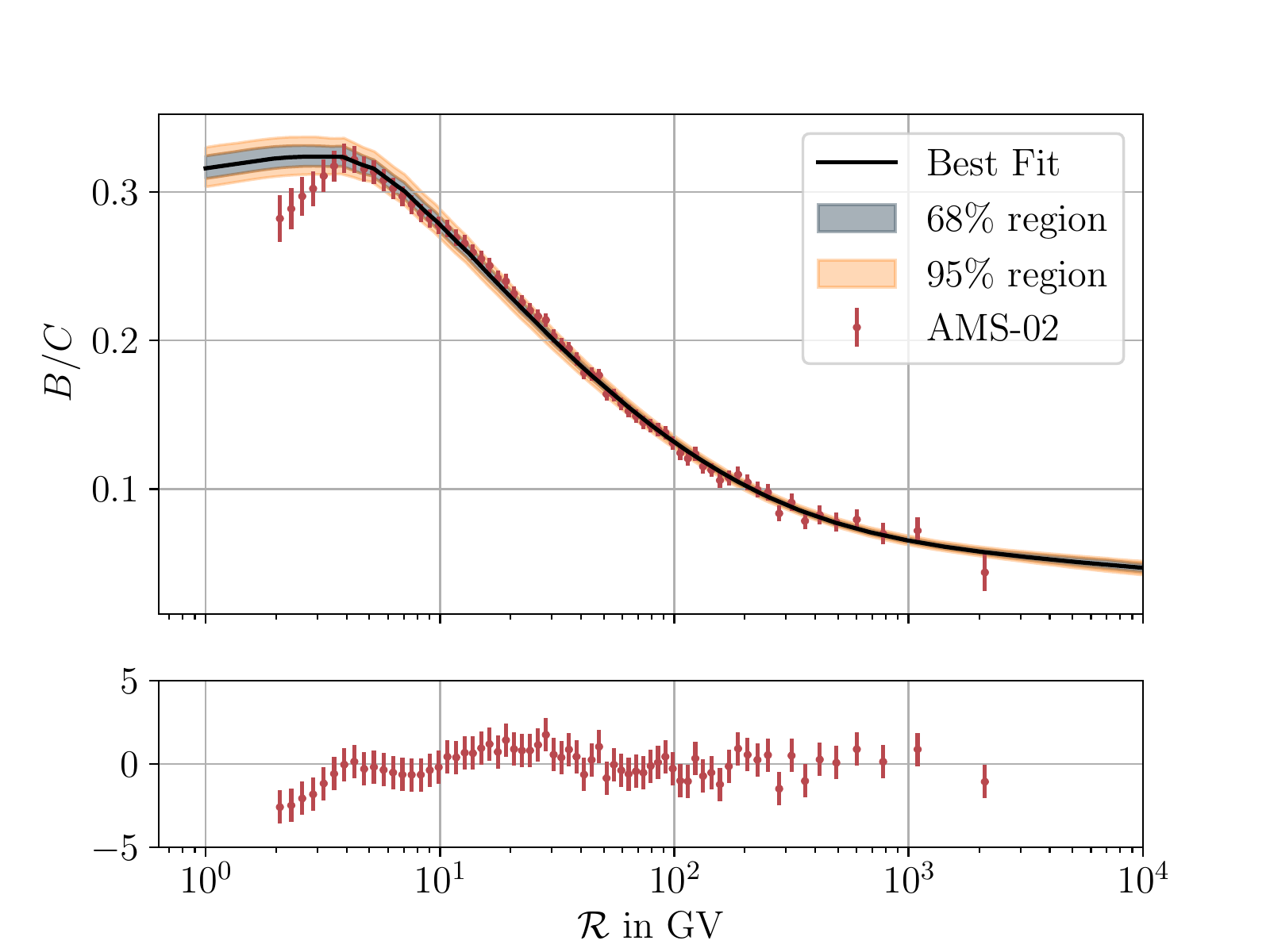} \includegraphics[width=0.49\textwidth]{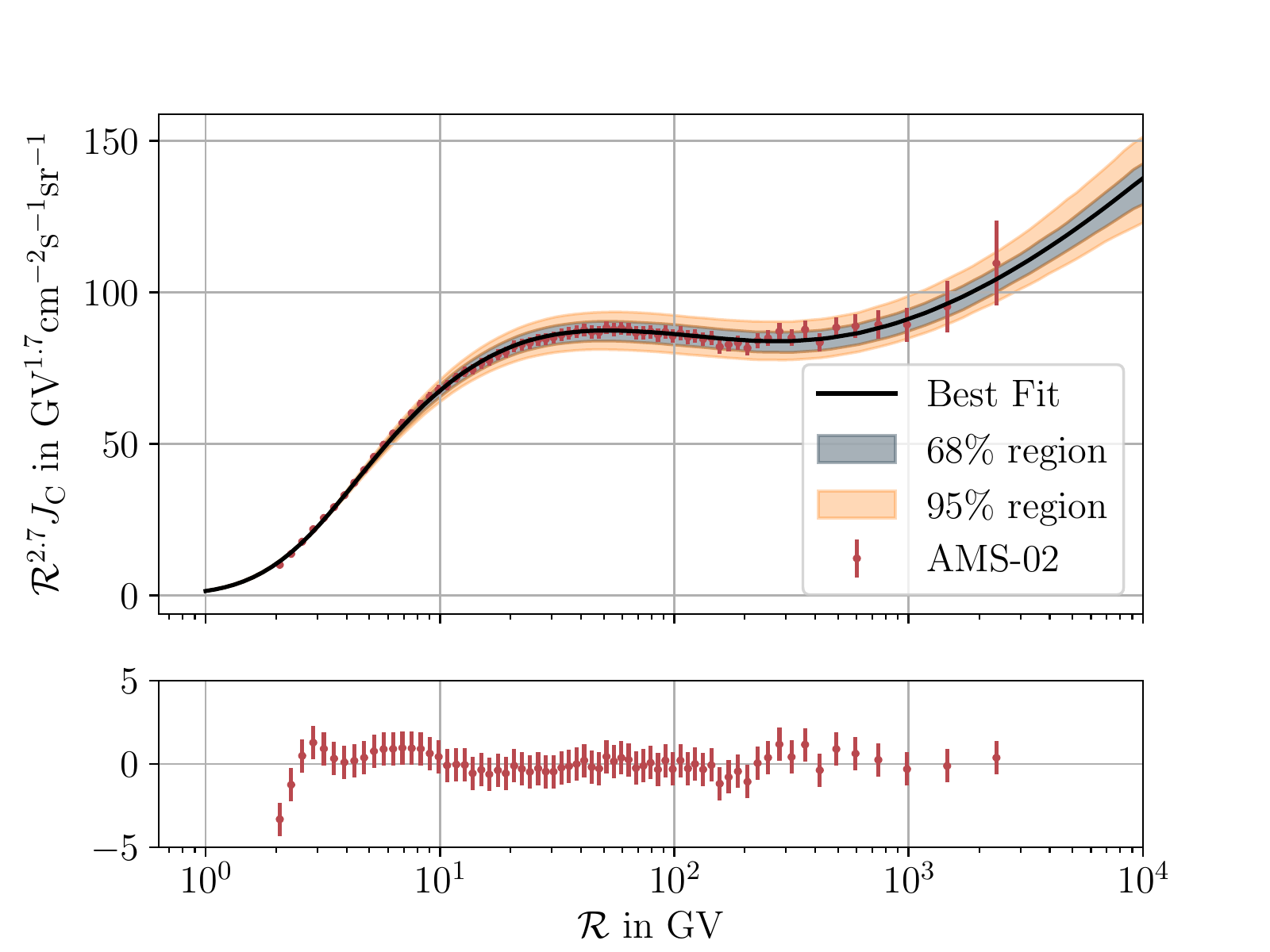} 
\caption{
Best-fit spectra for various GCR primaries and secondaries: protons (top left), helium (top right), electron (middle left), positrons (middle right), boron-to-carbon ratio (bottom left) and carbon (bottom right). 
The top panels show the GCR intensities, with the solid black line indicating the best-fit model and the grey and light orange bands showing the $68 \, \%$ and $95 \, \%$ uncertainty intervals. 
We have also overplotted the observations by AMS-02~\citep{AMS:2015tnn,AMS:2017seo,AMS:2019iwo,AMS:2019rhg,AMS:2018tbl}, DAMPE~\citep{DAMPE:2019gys,Alemanno:2021gpb}, IceTop~\citep{IceCube:2019hmk} and KASCACE~\citep{KASCADE:2005ynk}. 
The lower panels are pull plots. 
}
\label{fig:spectra}
\end{figure*}

Finally, our best-fit model predictions for the proton, helium, electron, positron and carbon intensities as well as boron-to-carbon ratio are shown in Figure~\ref{fig:spectra}. 
The proton and helium intensities are shown as a function of $E_{\text{kin}}/n$ as this is the relevant quantity for the production of diffuse emission. 
For the helium intensities, transforming the experimental data to match these units requires an assumption about the $^3\text{He}/^4\text{He}$ ratio.
We use the AMS-02 result $^3\text{He}/^4\text{He}\;(R)=0.1476\kl{R / 4\;\mathrm{GV}}^{-0.294}$ provided in \citet{Aguilar2021} and extrapolate this to higher energies.
We emphasize that this transformation is relevant only for the illustration in Figure~\ref{fig:spectra} and that all datasets are fitted in their respective measured units.

\begin{table}[!tbh]
\caption{$\chi^2$ of fits to local cosmic ray data.}
\small
\label{tbl2}
\begin{tabular}{l l c c}
\hline\hline
Experiment & Dataset & \# Data points & $\chi^2$ \\ \hline
AMS-02 & Protons&72 &22.0 \\
AMS-02 & Helium& 68&14.8 \\
AMS-02 & Carbon& 68&32.2 \\
AMS-02 & B/C& 67&54.1 \\
AMS-02 & Electrons& 72&32.8 \\
AMS-02 & Positrons& 72&131.2 \\
DAMPE & Protons& 17&8.2 \\
DAMPE & Helium& 23&5.9 \\
KASCADE & Protons& 20&31.7 \\
KASCADE & Helium& 14&94.4\\
IceTop & Protons& 19&52.3 \\
IceTop & Helium& 13&39.7 \\
\hline\hline
\end{tabular}
\end{table}

The best-fit model (black line) reproduces well the GCR data with an overall satisfactory goodness of fit. 
We have listed the $\chi^2$ values for the individual observables in Table~\ref{tbl2}. 
The pull distributions highlight some systematic deviations, for instance in the boron-to-carbon ratio below a few GV. 
This is due to our restriction to a pure diffusion model; allowing for advection or reacceleration would lead to a better fit at GV rigidities~\citep{1995ApJ...441..209H}. 

The $68 \, \%$ and $95 \,\%$ uncertainty bands are shown in grey and light orange, respectively. 
They are narrow where the data are sufficiently constraining and wider where the data are less constraining. 
An example for the latter case are proton and helium spectra beyond the energies where direct observations are available, that is beyond a few hundred TeV/n. 
Despite the energy-rescaling parameters that were allowed to float freely, there are still some discrepancies between IecTop and KASCADE measurements.
As the diffuse neutrino flux is mostly produced by proton and helium, this will be the dominant uncertainty from the GCR fit. 

% ----------------------------------------------------------------------------------------
% ----------------------------------------------------------------------------------------
\subsection{Diffuse Gamma-Ray Intensities}
In this section, we present the gamma ray intensities predicted by the \texttt{CRINGE} model. 
As discussed above, the diffuse predictions depend on a number of inputs beyond the parameters of the GCR models that we have fitted to local observations. 
These are the spatial distribution of GCR sources, the spatial distribution of atomic and molecular hydrogen as well as the cross-sections for the production of gamma-rays and neutrinos. 
In addition, the gamma-ray intensities depend on the choice of the ISRF. 
Given our ignorance of the true source distribution, gas distribution, cross-sections and ISRF, the choice induces another source of uncertainty beyond the uncertainty from the GCR fit. 

In the following, we have chosen a combination of these inputs as a default model. 
In particular, we have adopted the source distribution by \citet{Ferriere2001}, the \texttt{GALPROP} galactic gas maps as well as the \texttt{AAfrag} production cross-sections. 
The fiducial Inverse Compton flux is calculated assuming the \texttt{GALPROP} ISRF model.

We refer to this choice of parameters together with the best-fit parameters of the GCR model as the fiducial model.\footnote{This model is available at\\ \href{https://doi.org/10.5281/zenodo.7373010}{https://doi.org/10.5281/zenodo.7373010}\label{fn:zenodo}} 
For the different sources of uncertainties, we have computed the respective standard deviations separately. 
For the uncertainty stemming from a different choice of GCR parameters, we have determined half of the central $68 \, \%$ range of the posterior distribution of intensities. 
For the other sources of uncertainties, we have fixed the GCR parameters to their best-fit values and computed the standard deviation from the the set of diffuse fluxes obtained for different choices of the input, as listed in Secs.~\ref{sec:source_injection}, \ref{sec:gas_maps}, \ref{sec:cross-sections} and \ref{sec:ISRFs}. 
In the following figures, we have stacked these uncertainties into uncertainty bands. 

\begin{figure*}
\centering
\includegraphics[scale=0.5]{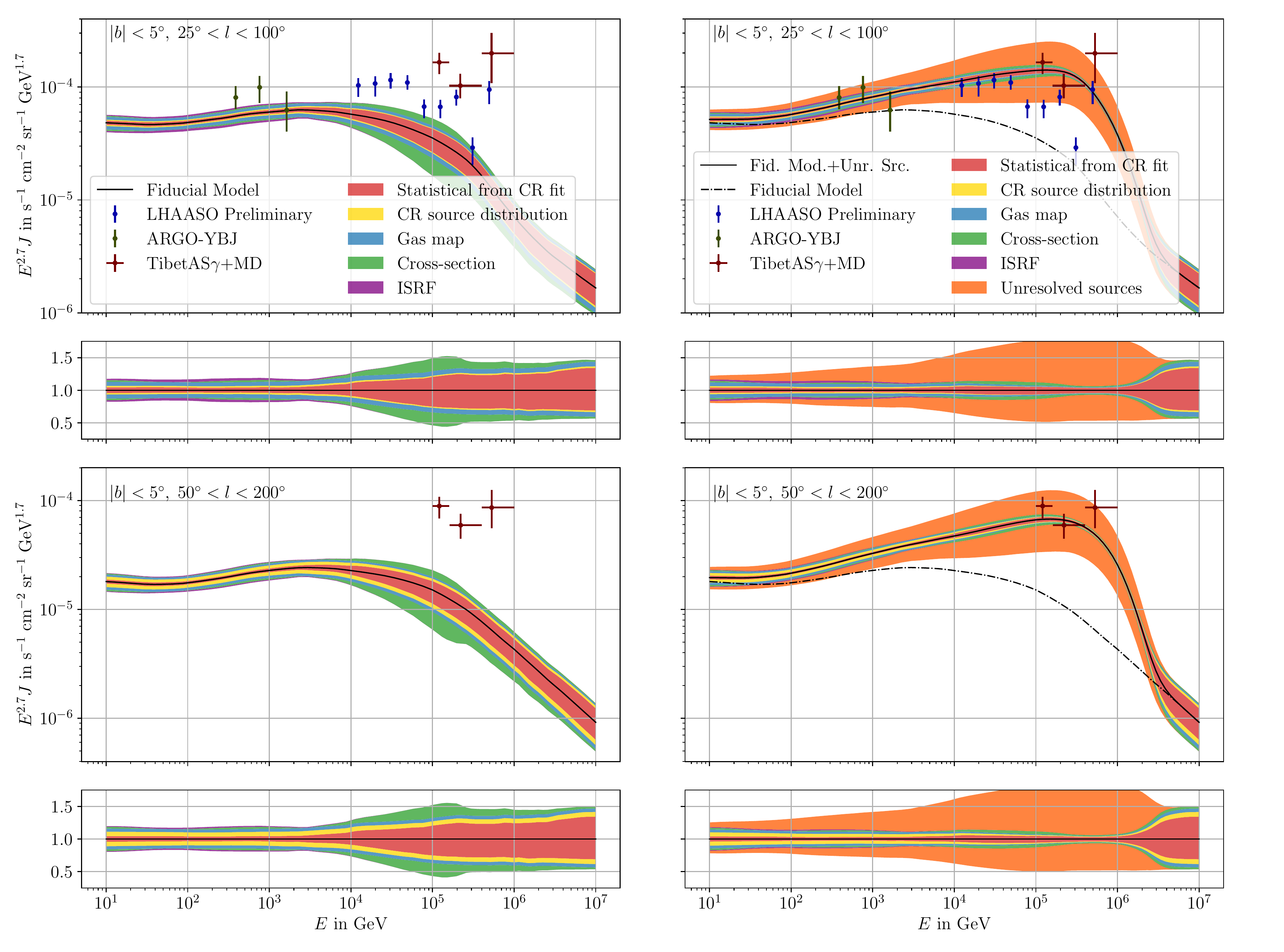}
\caption{
Gamma-ray intensities as a function of energy in the two windows $|b| < 5^{\circ}$, $25^{\circ} < l < 100^{\circ}$ (top row) and $|b| < 5^{\circ}$, $50^{\circ} < l < 200^{\circ}$ (bottom row). 
In the left column, we show our model prediction for the best-fit GCR parameters, combined with the \citet{Ferriere2001} source distribution, the \texttt{GALPROP} galactic gas maps, the \texttt{AAfrag} production cross-sections as well as the \texttt{GALPROP} ISRF. 
In the right column, we have also added the \citet{Vecchiotti2021a} model for the unresolved sources. 
The various uncertainties are indicated by the shaded bands. 
We also compare with the observations by Tibet AS$\gamma$+MD~\citep{TibetASgamma:2021tpz}, ARGO-YBJ~\citep{ARGO-YBJ:2015cpa} and LHAASO~\citep{Zhao2021}. 
The upper panels show the absolute intensities, the lower panels are normalized to the fiducial intensity. 
}
\label{fig:gamma_fiducial_gamma}
\end{figure*}

The fiducial model's intensity as well as the uncertainties around this intensity depend on direction in the sky. 
For gamma-ray intensities, we have adopted two sky windows for which observational data have been presented at TeV and PeV energies. 
In Figure~\ref{fig:gamma_fiducial_gamma}, we show the prediction of our model for the diffuse gamma-ray fluxes in the windows $|b| < 5^{\circ}$, $25^{\circ} < l < 100^{\circ}$ and $|b| < 5^{\circ}$, $50^{\circ} < l < 200^{\circ}$. 
Uncertainties due to the GCR parameters, the source distribution, the gas maps and the cross-sections are indicated by red, yellow, blue and green bands, respectively. 
Additionally, the uncertainty of the Inverse Compton intensity stemming from the uncertainty of the ISRF is indicated by the purple band.
We also take into account the expected intensity from unresolved sources following the model by \citet{Vecchiotti2021a}.  
The uncertainties related to the flux threshold varying between $0.01$ and $0.1 \, \Phi_{\text{Crab}}$ are shown by the orange bands. The default intensity of unresolved sources added to our fiducial diffuse model is the geometric mean of the intensities corresponding to the upper and lower end of that range.\footref{fn:zenodo} 

Without the inclusion of unresolved sources (left column of Figure~\ref{fig:gamma_fiducial_gamma}), our model spectrum is close to $E^{-2.7}$ for gamma-ray energies between $10 \, \text{GeV}$ and tens of TeV. 
Beyond a few tens of TeV, the spectrum softens due to the spectral breaks in the nuclear spectra at tens of TV and at the knee around $1 \, \text{PV}$, see  the nuclear spectra in Figure~\ref{fig:spectra}. 
As far as the uncertainties are concerned, below a TeV, the uncertainty from the gas maps is dominating, but the total uncertainty remains below $20 \, \%$. 
Above a TeV, the uncertainties from the GCR model and from the cross-sections grow; individually they can be as large as $35 \, \%$ and $20 \, \%$, respectively. 
The other uncertainties related to the choice of gas maps and cosmic ray source distribution are independent of energy and remain below $10 \, \%$ in all regions in the sky, respectively. 
Even within these uncertainties, our model without unresolved sources cannot account for the data by LHAASO~\citep{Zhao2021} and Tibet AS$\gamma$+MD~\citep{TibetASgamma:2021tpz}, in neither of the sky windows. 

The inclusion of the unresolved sources significantly enhances the intensities overall and leads to a much harder, close to $E^{-2.3}$ spectrum for gamma-ray energies between a few hundreds of GeV and a few hundreds of TeV. 
The right column of Figure~\ref{fig:gamma_fiducial_gamma} shows that the gamma-ray intensities are now in much better agreement with the data by LHAASO~\citep{Zhao2021} and Tibet AS$\gamma$+MD~\citep{TibetASgamma:2021tpz}, in both sky windows. 
However, the uncertainty in the prediction of the intensity from unresolved sources is sizeable: Between $\sim 100 \, \text{GeV}$ and a few PeV, it is dominating the total uncertainty and can be as large as a factor $2$.

% ----------------------------------------------------------------------------------------
% ----------------------------------------------------------------------------------------
\subsection{Diffuse Neutrino Intensities}
\label{sec:diffuse_neutrino_intensities}

\begin{figure*}
\centering
\includegraphics[scale=0.5]{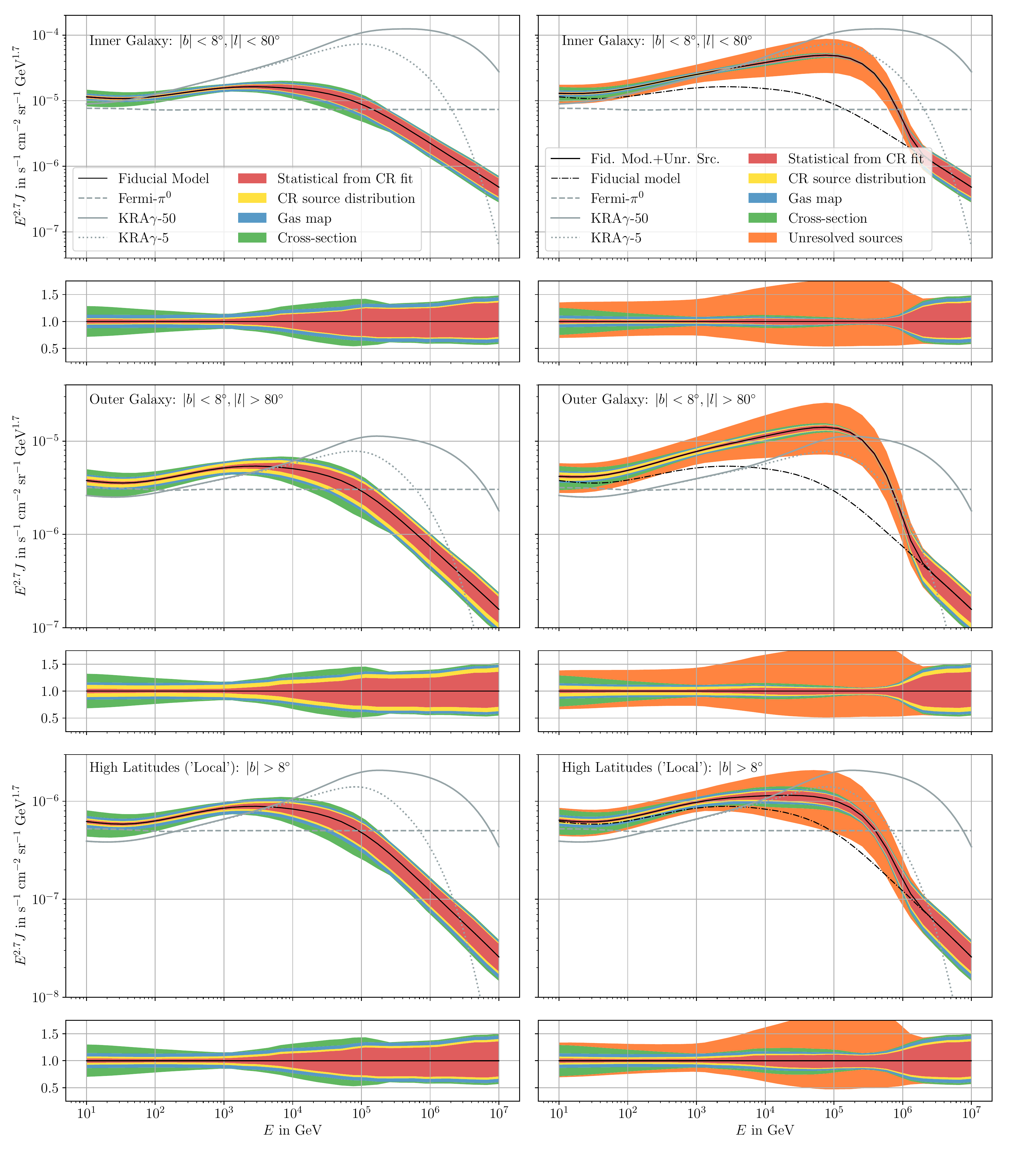}
\caption{
Neutrino intensity as a function of neutrino energy for the inner Galaxy window (top panels), outer Galaxy window (middle panels) and high-latitude window (bottom panels). 
In the left column, only truly diffuse emission is shown, in the right column, unresolved sources are included as well.
In each plot, we show our model prediction for the best-fit GCR parameters, combined with the \citet{Ferriere2001} source distribution, the \texttt{GALPROP} galactic gas maps as well as the \texttt{AAfrag} production cross-sections. 
The Fermi-$\pi^0$ model is indicated by the dashed grey line, the KRA$\gamma$-5 and KRA$\gamma$-50 models by the solid and dotted grey lines. 
The coloured bands in the top panel indicate the uncertainties due to the GCR parameters (red band), the source distribution (yellow band), the gas maps (blue bands) and the cross-sections (green band).
In the lower panels, the uncertainty bands are presented after normalization to the fiducial model. 
}
\label{fig:nu_fiducial}
\end{figure*}

For neutrinos, we have chosen three regions of the sky over which the diffuse intensities are averaged: 
An inner Galaxy window ($|b| < 8^{\circ}$, $|l| < 80^{\circ}$), an outer Galaxy window ($|b| < 8^{\circ}$, $|l| > 80^{\circ}$) and a high-latitude window ($|b| > 8^{\circ}$). 
These are canonical choices in models of galactic diffuse emission. 
In Figure~\ref{fig:nu_fiducial}, the fiducial model intensities are again shown by the black, solid lines and the uncertainties are indicated by bands of different colours: GCR parameters (red), source distributions (yellow), gas maps (blue) and cross-sections (green). 
In the lower panel, the uncertainty bands are shown after normalization to the fiducial model intensity. 
We also indicate the predictions of other models, that is the Fermi-$\pi^0$ model~\citep[dashed grey line]{Fermi-LAT:2012edv}, the KRA$\gamma$-5 model~\citep[dotted grey line]{Gaggero:2014xla} as well as the KRA$\gamma$-50 model~\citep[solid grey line]{Gaggero:2014xla}. 
All neutrino intensities shown are per-flavor intensities. We derive these from the all-flavor intensity under the assumption that neutrino oscillations lead to a $1:1:1$ flavor ratio at Earth \citep{Gaisser2016}. 
For the inner Galaxy window and without unresolved sources (top left panel of Figure~\ref{fig:nu_fiducial}), our model spectrum is close to $E^{-2.7}$ for neutrino energies between $10 \, \text{GeV}$ and tens of TeV. 
Beyond a TeV, the spectrum softens due to the spectral breaks in the nuclear spectra. 
Below a TeV, the uncertainty is dominated by the cross-section uncertainty which can be as large as $20 \, \%$. 
At $\sim 1 \, \text{TeV}$, the cross-section uncertainties are smallest, and grow again for higher energies. 
Also the uncertainties from the GCR model grow significantly beyond a TeV and reach about $40 \, \%$ at $10 \, \text{PeV}$. 
The other uncertainties related to the choice of gas maps and cosmic ray source distribution are again independent of energy and remain below $10 \, \%$ in all regions in the sky, respectively. 

Comparing our spectrum in the inner Galaxy window with the largely featureless, $E^{-2.7}$ spectrum of the Fermi-$\pi^0$ model, we find our model to give slightly harder spectra below a few TeV, due to the break in the nuclei spectra at $\sim 300 \, \text{GV}$ that had not been considered in the Fermi-$\pi^0$ model. 
Above tens of TeV, though, the Fermi-$\pi^0$ model is clearly harder due to the unbroken power law extrapolation of the spectra from GeV to PeV energies. 
We judge that extrapolation to not be well justified in light of data at the knee and recent data just below the knee. 
Below a few hundred TeV and a few PeV, the predictions from the KRA$\gamma$-5 and KRA$\gamma$-50 models, respectively, are significantly harder than ours. 
This is of course in part due to the harder spectral index exhibited by these models in the inner Galaxy, but also due to the choice of the cross-sections from \citet{Kamae2006}, which as shown in Figure~\ref{fig:xsec_profiles}, leads to systematically harder spectra than the \texttt{AAfrag} parametrization adopted in our fiducial model. 
Already at $100 \, \text{TeV}$, both models overpredict our neutrino intensity by roughly an order of magnitude. 
At a few PeV, this difference has grown to almost two orders of magnitude in the case of the KRA$\gamma$-50 model. 
Note that spectra are generally softer above a few hundreds of TeV (a few PeV) for KRA$\gamma$-5 (KRA$\gamma$-50) due to the assumed exponential cut-off. 

The inclusion of unresolved sources in the inner Galaxy window (top right panel of Figure~\ref{fig:nu_fiducial}) leads to a much enhanced neutrino intensity below $\sim 1 \, \text{PeV}$. 
Our prediction is much closer to the prediction from the KRA$\gamma$-5 model, even though the origin of the hard spectrum is very different. 

\begin{figure*}[tbh]
\centering
\includegraphics{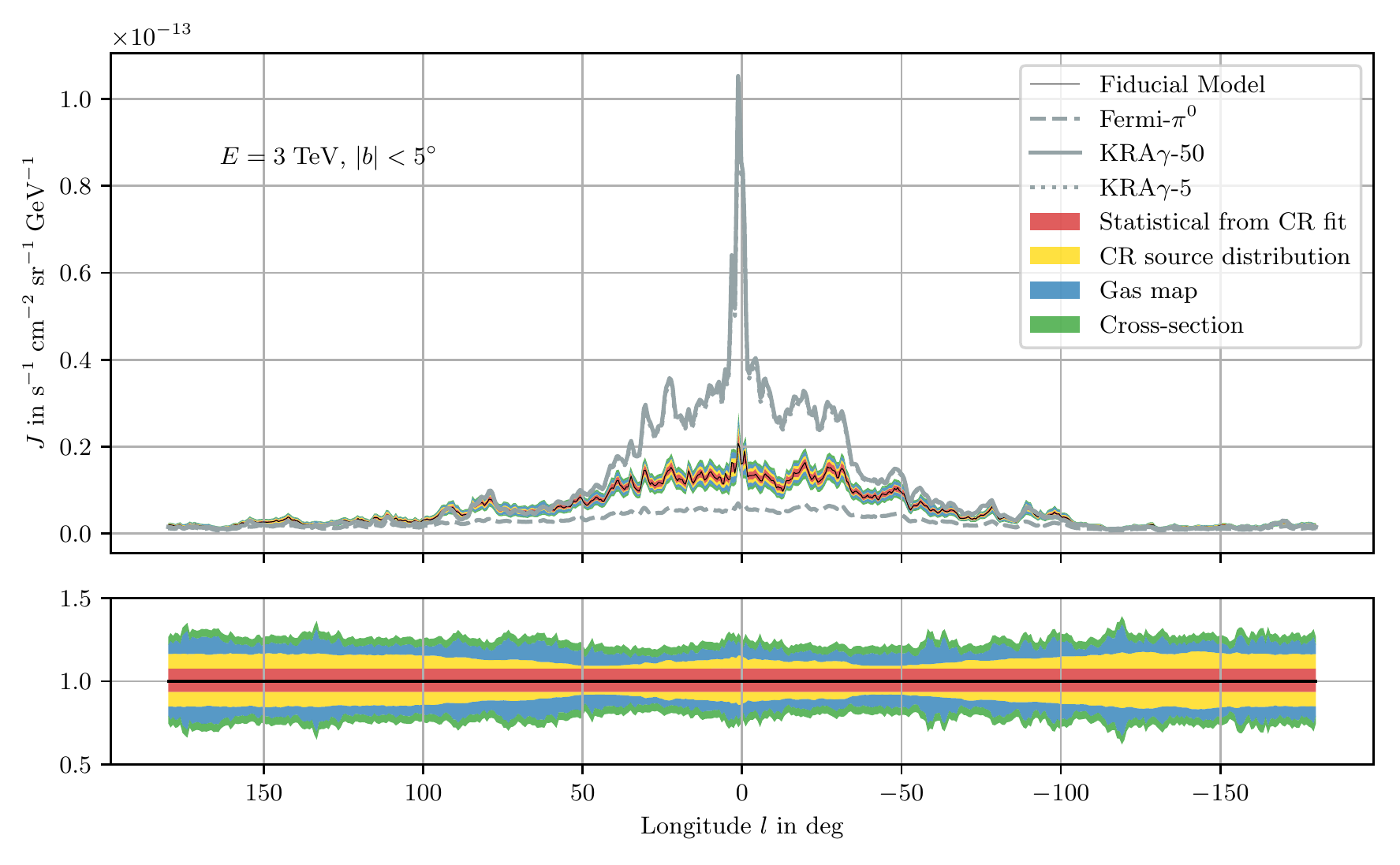}
\caption{
Neutrino intensity as a function of galactic longitude for a neutrino energy of $3 \, \text{TeV}$. 
We show our model prediction for the best-fit GCR parameters, combined with the \citet{Ferriere2001} source distribution, the \texttt{GALPROP} galactic gas maps as well as the \texttt{AAfrag} production cross-sections. 
The Fermi-$\pi^0$ model is indicated by the dashed grey line, the KRA$\gamma$-5 and KRA$\gamma$-50 models by the solid and dotted grey lines. 
In the lower panel, the
uncertainty bands are presented after normalization to the fiducial model.
}
\label{fig:gamma_fiducial_profile}
\end{figure*}

For the outer Galaxy window (middle row of Figure~\ref{fig:nu_fiducial}), the intensities are overall smaller by about a factor two to three, but the spectral shapes are rather similar. 
This is to be expected given that we assume no spatial variation of the diffusion coefficient or the source spectra throughout the Galaxy. 
Noticeable is, however, an increased uncertainty from the source distribution
(compare the yellow bands in the upper and middle rows of Figure~\ref{fig:nu_fiducial}). 
At first this might seem surprising, given that in absolute terms, the source distributions differ less in the outer Galaxy than in the inner Galaxy, see Figure~\ref{fig:sd_radial_profiles}.
However, this can be explained by the fact that, for the same local source density, the distributions featuring a lower source density in the inner Galaxy lead to a lower cosmic ray flux at Earth. 
As we however normalize the local flux of our models such that local measurements are reproduced, we correspondingly scale up the fluxes for lower central source densities. This decreases the resulting uncertainty in the galactic center region and increases it towards larger galactocentric radii.

For the case without unresolved sources (middle left panel of Figure~\ref{fig:nu_fiducial}), the comparison with the Fermi-$\pi^0$ model in this sky window shows that both models are compatible between $10$ and $100 \, \text{GeV}$, but at larger energies show spectral differences similar to those in the inner Galaxy window. 
The KRA$\gamma$ models are at the lower envelope of the uncertainty band below a few TeV, but start exceeding the upper end of our uncertainty band above $\sim 10 \, \text{TeV}$. 
Again, the disagreement can be up to two orders of magnitude at $1 \, \text{PeV}$. 

For the case where unresolved sources are taken into account (middle right panel of Figure~\ref{fig:nu_fiducial}), the intensities are again significantly harder. 
However, in this case the prediction is larger even than that of the KRA$\gamma$-5 model for all energies and larger than that of the KRA$\gamma$-50 model for neutrino energies smaller than a few hundred TeV. 

The situation is somewhat similar for high galactic latitudes, as shown in the lower row of Figure~\ref{fig:nu_fiducial}, albeit at an overall normalization reduced by a factor $\sim 6$ with respect to the outer Galaxy window. 
The uncertainties from the source distribution are in between those for the inner and outer Galaxy windows, as could be expected. 

In Figure~\ref{fig:gamma_fiducial_profile}, we also show profiles of the neutrino intensity in galactic longitude at an energy of $3 \, \text{TeV}$. 
We have chosen an energy at which all sources of uncertainties contribute $\sim 10 \, \%$ to the overall uncertainty. 
The prediction from our fiducial model with the \citet{Ferriere2001} source distribution, the \texttt{GALPROP} gas maps and the \texttt{AAfrag} cross-sections is again shown by the black line and uncertainties around that are denoted by the bands in the same fashion as in Figure~\ref{fig:nu_fiducial}. 
We also compare with the longitudinal profiles of the Fermi-$\pi^0$ and the KRA$\gamma$ models. 
The linear ordinate axis highlights the large differences between those models and our model, in particular for the inner Galaxy. 
While the uncertainty from the GCR model and the cross-section uncertainties are independent of longitude, the uncertainties from the source distribution and gas maps, of course, depend on longitude. 
Specifically, the CR source distribution uncertainty is largest in the outer Galaxy, has a minimum for $|l| \simeq 45^{\circ}$, where on average the emission is produced at galacto-centric radii similar to the solar radius, and increases again towards the galactic center.
The uncertainties from the gas maps instead are largest
towards $|l| \simeq 45^{\circ}$.
This can be interpreted as a generic model uncertainty of the map reconstruction: Towards this direction the gradients in the velocity field are rather small such that the distance uncertainty is largest.

% ----------------------------------------------------------------------------------------
% ----------------------------------------------------------------------------------------
% ----------------------------------------------------------------------------------------
\section{Discussion\label{sec:discussion}}

\begin{figure}
\centering
\includegraphics[width=0.49\textwidth]{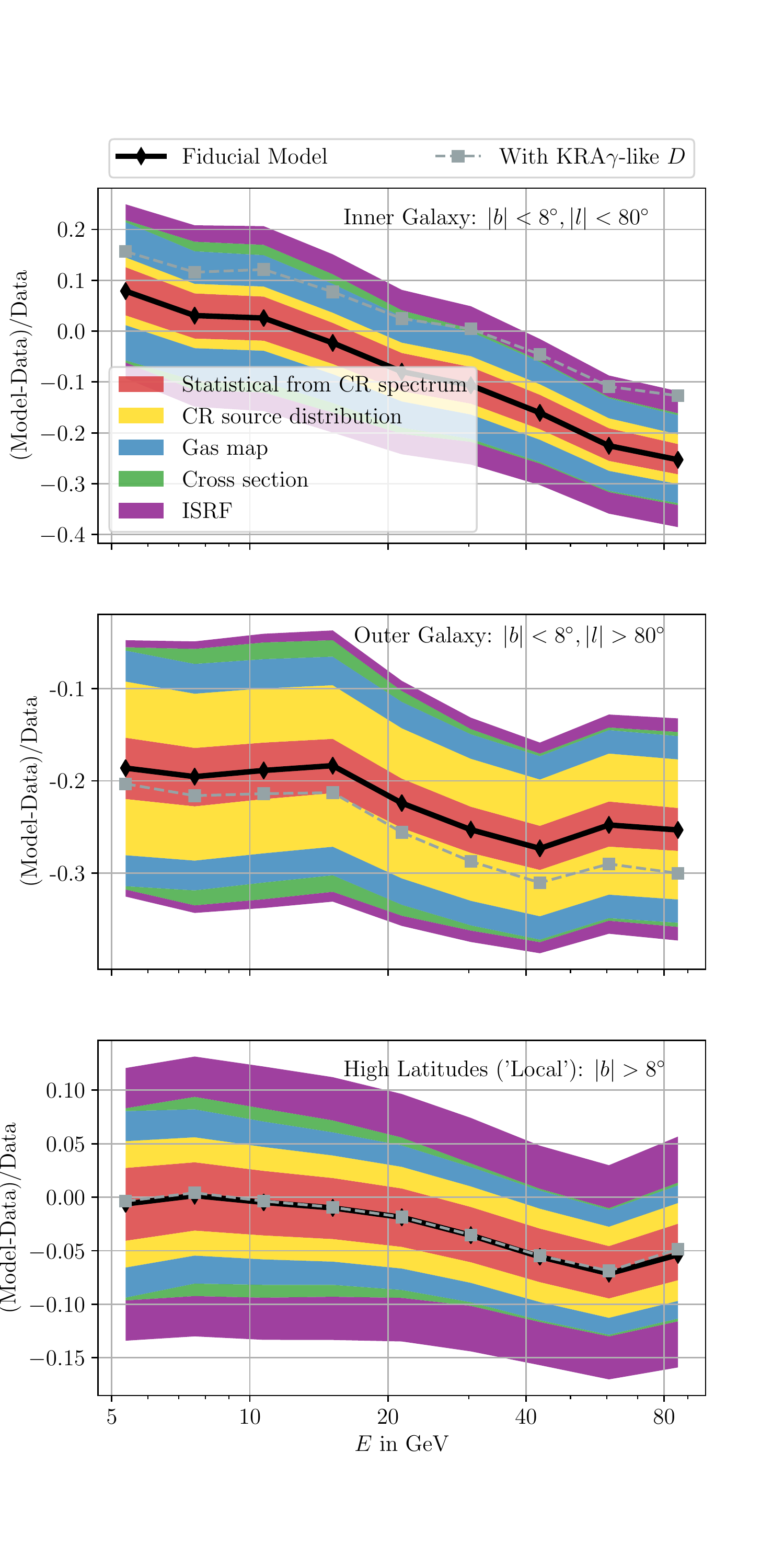}
\caption{
Relative count differences to P8R3 ULTRACLEANVETO \textit{Fermi}-LAT data for for our fiducial model combining the best-fit GCR parameters with the \citet{Ferriere2001} source distribution, the \texttt{GALPROP} galactic gas maps as well as the \texttt{AAfrag} production cross-sections. Also shown are the relative count difference for a model that has all parameters equal to this fiducial model apart from the spectral indices of the diffusion coefficient, which depend on galactocentric radius as described in eq.~\eqref{eq:kragammadiffusion} with $A_i=0.035\;\mathrm{kpc}^{-1}$. The $B_i$ are chosen such that $\delta_i(r=r_\odot)$ agree with the values from Table~\ref{tbl1}.
}
\label{fig:GeV_gammas}
\end{figure}

In this paper, we have predicted the high-energy diffuse neutrino intensity from the Galaxy, based on a model of GCRs that fits measured cosmic ray data between GV and $100 \, \text{PV}$ rigidities. 
We have also predicted diffuse gamma-ray fluxes in the TeV to PeV energy range and compared to results from ARGO-YBJ, Tibet AS-$\gamma$+MD as well as LHAASO. 
It is, however, natural to ask whether our models agree also with GeV gamma-rays as for instance measured by \textit{Fermi}-LAT~\citep{Fermi-LAT:2009ihh}. 
Such a comparison is of course made difficult by the various anomalies discussed above: the hardening of diffusive GeV emission towards the inner Galaxy, the higher than predicted intensities in the inner Galaxy as well as the galactic center excess. 
Admittedly, the model framework considered here does not alleviate any of these anomalies. 
For instance, we show in Figure~\ref{fig:GeV_gammas}, the relative differences between the measured \textit{Fermi}-LAT counts\footnote{We are grateful to Markus Ackermann for providing us with the counts maps and instrument response functions.} and the prediction of our fiducial model for three different sky regions also shown in Figure~\ref{fig:nu_fiducial}. 
In the inner Galaxy window (top panel), it can be seen that while the model is in agreement with the GeV gamma-ray data within uncertainties except at energies above $40 \, \text{GeV}$, the observed gamma-ray fluxes are harder than the model fluxes. 
In addition to the central Galaxy, there is a significant discrepancy between GeV gamma-ray data and our model predictions in the outer Galaxy window, where we underproduce the data by some $20 \, \%$. We note that this issue has received relatively little attention yet and therefore awaits further clarification. 
Finally, we find our model to largely reproduce the data at high latitudes, unlike some of the previous studies~\citep{Orlando2018}. 

We have also explored the framework of KRA$\gamma$-like models which aim for solving the puzzle of the observed spectral hardening in \textit{Fermi}-LAT data towards the galactic center by modifying the diffusion constant in the inner Galaxy as
\begin{equation} \label{eq:kragammadiffusion}
    \delta_i(r)=
    \begin{cases}
        A_ir+B_i & r<11\;\mathrm{kpc}\\
        A_i\times11\;\mathrm{kpc} + B_i & r\geq 11\;\mathrm{kpc}\\
    \end{cases},
\end{equation}
leading to harder spectra from that region. 
However, we have found that such modifications can only partly solve the hardening issue. 
This is illustrated in the upper panel of Figure~\ref{fig:GeV_gammas}. There, we have superimposed the prediction for a KRA$\gamma$-like model that is equal to our fiducial model apart from the diffusion coefficient, which depends on galactocentric radius as described in eq.~\eqref{eq:kragammadiffusion} with $A_i=0.035\;\mathrm{kpc}^{-1}$. The $B_i$ are chosen such that $\delta_i(r=r_\odot)$ agree with the values from Table~\ref{tbl1}. 
The residuals of this model show a trend rather similar to our fiducial model with predicted spectra that feature a larger flux normalization but remain too soft. This is evidence that a significant part of the enhancement and spectral hardening of the KRA$\gamma$ models with respect to the Fermi-$\pi^0$ model (which assumes homogeneous diffusion) is only due to the hardening break in the diffusion coefficient at around $300 \, \text{GV}$. 

In the outer Galaxy window, the KRA$\gamma$-like model even features slightly larger residuals with the \textit{Fermi}-LAT data than our fiducial model. At high latitudes, where local emission dominates, the two models coincide as expected.

Another interesting outcome of our global fit of the GCR model is the spectral position of the knee. 
The break in the predicted proton spectrum, for instance occurs at a few hundred TeV which is somewhat lower than traditionally considered. 
Note that this energy might in fact be compatible with claims by ARGO-YBJ~\cite{ARGO-YBJ:2015isx}. 
Returning to Figure~\ref{fig:spectra}, we note that the low value for the break rigidity $\mathcal{R}_{45}$ has been driven by the KASCADE helium data. 
In a sense, the low $\mathcal{R}_{45}$ is not surprising, but a consequence of the 300 GV break which leads to the dominance of helium even before the proton knee~\citep{OCDrury:2017rgu}.
The only way, in fact, to get a more pronounced peak at higher rigidities would be to change the energy scale corrections with respect to what our fit determined. 
This could be possible if there was another hardening break between 10 TV and 1 PeV.
We note that there have been indications from DAMPE~\citep{DAMPE_ICHEP2022}, yet, we have not added these data preliminary yet.

% ----------------------------------------------------------------------------------------
% ----------------------------------------------------------------------------------------
% ----------------------------------------------------------------------------------------
\section{Summary and conclusion\label{sec:conclusion}}

We have presented \texttt{CRINGE}, a new model for the diffuse emission of high-energy gamma-rays and neutrinos from the Galaxy. 
For the transport of GCRs, we have adopted a simple diffusion model with homogeneous diffusion coefficient, but we have allowed for a number of breaks in the rigidity-dependence. 
We have determined the free parameters of our model by fitting to locally measured spectra of proton, helium, electrons and positrons as well as to the boron-to-carbon ratio. 
Adopting an MCMC method, we have determined the uncertainties of the fitted parameters and provided uncertainty bands for the predicted GCR intensities and the boron-to-carbon ratio. 
Our GCR model successfully describes these data in the energy range between $1 \, \text{GeV}$ and $100 \, \text{PeV}$. 

Combining the best-fit GCR parameters with a fiducial choice for the source distribution, gas maps, cross-sections and photon backgrounds, we have computed the diffuse emission of high-energy gamma-rays and neutrinos at energies between $10 \, \text{GeV}$ and $10 \, \text{PeV}$. 
We have also estimated the uncertainties due to the possibility of alternative choices of inputs. 
In addition, we have allowed for the presence of unresolved, pulsar-powered sources of high-energy gamma-rays and neutrinos. 
Comparing our results with the intensity of high-energy gamma-rays observed by Argo-YBJ, Tibet AS$\gamma$+MD and LHAASO, we have found very good agreement for the case that unresolved sources are taken into account.
For our neutrino predictions, we have compared with the Fermi-$\pi^0$ and the KRA$\gamma$ models that have previously been employed in experimental searches. 
Without the unresolved sources, our predictions are usually lower than the KRA$\gamma$ models for $E \gtrsim 1 \, \text{TeV}$ and higher than the Fermi-$\pi^0$ model for $E \lesssim 100 \, \text{TeV}$. 
Taking the unresolved sources into account, our predictions are mostly comparable to or even higher than the KRA$\gamma$ models. 

Future studies of the galactic diffuse neutrino flux by IceCube and KM3Net will be able to employ our model predictions as spatio-spectral templates. 
Such searches present an important test of the model of GCRs. 
The possible observation of or bounds on the diffuse galactic flux of high-energy neutrinos will present important constraints on models of acceleration and transport of GCRs at TeV and PeV energies. 
In addition, we can hope to gain some insights on the presence of unresolved hadronic sources. 
% ----------------------------------------------------------------------------------------
% ----------------------------------------------------------------------------------------
\begin{acknowledgments}
We are grateful to Markus Ackermann for providing us with the \textit{Fermi}-LAT counts maps and instrument response functions. We also thank Mischa Breuhaus for providing his implementation of the calculation of gamma-ray absorption in the Milky Way. Daniele Gaggero, Pedro de la Torre Luque and Ottavio Fornieri are gratefully acknowledged for help with the \texttt{DRAGON} code. We also thank Carmelo Evoli for advice on the \texttt{HERMES} code. We acknowledge use of the \texttt{HEALPix} package~\citep{Gorski2005}. G.S. acknowledges membership in the International Max Planck Research School for Astronomy and Cosmic Physics at the University of Heidelberg (IMPRS-HD).
\end{acknowledgments}
% ----------------------------------------------------------------------------------------
% ----------------------------------------------------------------------------------------
% ----------------------------------------------------------------------------------------
\bibliography{library}
\begin{appendix}
\section{Comparison to \textit{Fermi}-LAT Data}
In the following, we provide some details on the preparation of \textit{Fermi}-LAT data that we use in Sec.~\ref{sec:discussion}. 
In order to compare a diffuse model to pixelized \textit{Fermi}-LAT count maps, we have used forward folding.  
For a given gamma-ray intensity $J$, the expected counts per solid angle and reconstructed energy are calculated as
\begin{equation}
\mathcal{C}(E_{reco},\boldsymbol{\theta}_{reco})=\int \dd E \int \dd \Omega\,\Phi_\gamma(E,\boldsymbol{\theta}) \mathcal{E}(E,\boldsymbol{\theta})\mathrm{PSF}(|\boldsymbol{\theta}-\boldsymbol{\theta}_{\text{reco}}|,E) \mathrm{ED}(E_{\text{reco}}|E).
\end{equation} 
with the instrument response functions (IRFs) exposure $\mathcal{E}$, energy-dependent point spread function $\mathrm{PSF}$ and the energy dispersion $\mathrm{ED}$.
From this, the expected counts in pixel $i$ of energy bin $k$, $\mu_{ik}$, follow as
\begin{equation}
\mu_{ik}=\int^{E_{\text{max},k}}_{E_{\text{min},k}} \dd E_{\text{reco}} \int_{\Omega_i}d\Omega\,\mathcal{C}(E_{\text{reco}},\boldsymbol{\theta}_{\text{reco}})
\end{equation}
This quantity is then compared directly to the \textit{Fermi}-LAT data counts to obtain the results presented in Figure~\ref{fig:GeV_gammas}.

The \textit{Fermi}-LAT dataset used in this work consists of 8 years of Pass 8 P8R3 ULTRACLEANVETO data from the same time window as used for the construction of the 4FGL catalogue \citep{Fermi2019} together with the corresponding IRFs. Pass 8 P8R3 is the latest \textit{Fermi}-LAT data release featuring the current up-to-date processing and event reconstruction \cite{Bruel2018}. The ULTRACLEANVETO event selection features the lowest instrumental background and is recommended for the study of large scale structures \citep{Bruel2018}. The dataset used in this work features events binned into 22 logarithmic energy bins ranging from $1.6\;\mathrm{GeV}$ to $2.32\;\mathrm{TeV}$. Besides that, events are separated into four classes according to the quality of their angular reconstruction. These are labelled PSF0-PSF3, with a larger number representing better reconstruction quality, and each contain about the same number of events, but feature a separate PSF.\footnote{\href{https://fermi.gsfc.nasa.gov/ssc/data/analysis/documentation/Cicerone/Cicerone_LAT_IRFs/}{https://fermi.gsfc.nasa.gov/ssc/data/analysis/documentation/Cicerone/Cicerone\_LAT\_IRFs/}\label{fn:lat_irf}}

The total gamma-ray flux measured by \textit{Fermi}-LAT consists of course not only of galactic diffuse emission, but also of emission from both point-like and extended sources as well as isotropic extragalactic emission and the remaining instrumental backgrounds. For the latter, a recommended model for the P8R3 ULTRACLEANVETO selection is provided publicly\footnote{\href{https://fermi.gsfc.nasa.gov/ssc/data/access/lat/BackgroundModels.html}{https://fermi.gsfc.nasa.gov/ssc/data/access/lat/BackgroundModels.html}} and included as an additional flux component in the model. Also, in addition to the hadronic diffuse emission and the Inverse Compton component discussed in the main text, we also include the contribution from bremsstrahlung of cosmic ray leptons. This component can be safely neglected above $20\,\mathrm{GeV}$ due to its steeply falling spectrum, but contributes at the level of up to $5\,\%$ at energies of a few GeV. With this, the total intensity that is compared to data is
\begin{equation}
J=J_{\mathrm{had}}+J_{\mathrm{IC}}+J_{\mathrm{brems}}+J_{\mathrm{isotropic}}+J_{\mathrm{instrument}}
\end{equation}
Note that this does not include the intensity from either resolved or unresolved sources. While we do not consider unresolved sources in this analysis, we mask out known sources to stay as independent of the details of source emission models as possible. This includes all 5064 sources included in the 4FGL, of which 75 are spatially extended \citep{Fermi2019} as well as two large scale features in the galactic gamma-ray sky that can not be reproduced with the modelling approach pursued here, the Fermi bubbles \citep{Su:2010qj} and the north polar spur (NPS) \citep{Fermi-LAT:2016zaq,Casandjian2009}. For the Fermi bubbles, a publicly available spatial template\footnote{\href{https://dspace.mit.edu/handle/1721.1/105492}{https://dspace.mit.edu/handle/1721.1/105492}} similar to that shown in \citet{Fermi-LAT:2016zaq} is used, while the mask for the NPS is constructed following the template shown in \citet{Fermi-LAT:2016zaq}. For the sources from the 4FGL, a circular mask with radius $\sigma_{\text{mask}}$ is assumed. For the spatially extended sources, the mask size is not $\sigma_{\text{mask}}$, but rather the larger $\sigma_{ext}=\sqrt{\sigma_{\text{mask}}^2+\sigma^2_{\text{src}}}$, with $\sigma_{\text{src}}$ the radius of the source. $\sigma_{\text{mask}}$ must of course be chosen large enough to mask out a large fraction of the source emission, but also needs to be small enough to still leave a significant fraction of pixels unmasked, in particular towards the galactic center where the source density is highest. As a compromise between these considerations, a value of $\sigma_{\text{mask}}=0.5^\circ$ is chosen, and the data considered in this work are restricted to the PSF3 and PSF2 classes at energies $E>5\;\mathrm{GeV}$. This ensures that all sources are masked out at at least $2.3\;\times$ the $68\,\%$ PSF containment angle. For simplicity, and to be even more restricting at higher energies, $\sigma_{\text{mask}}$ is fixed independently of gamma-ray energy. As a consequence of only considering the \textit{Fermi}-LAT data above $5\;\mathrm{GeV}$, where the energy resolution is much better than at lower energies\footref{fn:lat_irf} \citep[e.g.][]{Atwood2009}, energy dispersion is neglected. Given this energy threshold and the rather wide logarithmic energy binning of the dataset used here with $E_{\text{max},i} / E_{\text{min},i}=\sqrt{2}$, the systematic error introduced by this simplification was found to be below $5\,\%$ in all cases. On the high-energy end, we restrict our analysis to gamma-ray energies below $100\;\mathrm{GeV}$. This cut-off ensures that the Poisson uncertainty on the data counts also remains below $5\,\%$ in all cases.

\section{Input variations for best-fit GCR parameters}
The quantification of the uncertainties of galactic diffuse emission stemming from the choice of gas distributions, hadronic production cross-sections, cosmic ray source distribution and interstellar radiation field as described in Sec.~\ref{sec:diffuse_neutrino_intensities} are based on discrete variations of the respective model inputs assuming the best-fit GCR parameters. In the following, we show the fluxes for each of these discrete variations to illuminate the shape of the uncertainty bands displayed in Sec.~\ref{sec:results}.
\subsection{Hadronic Production Cross-sections}

\begin{figure*}[h]
\centering
\includegraphics[scale=0.495]{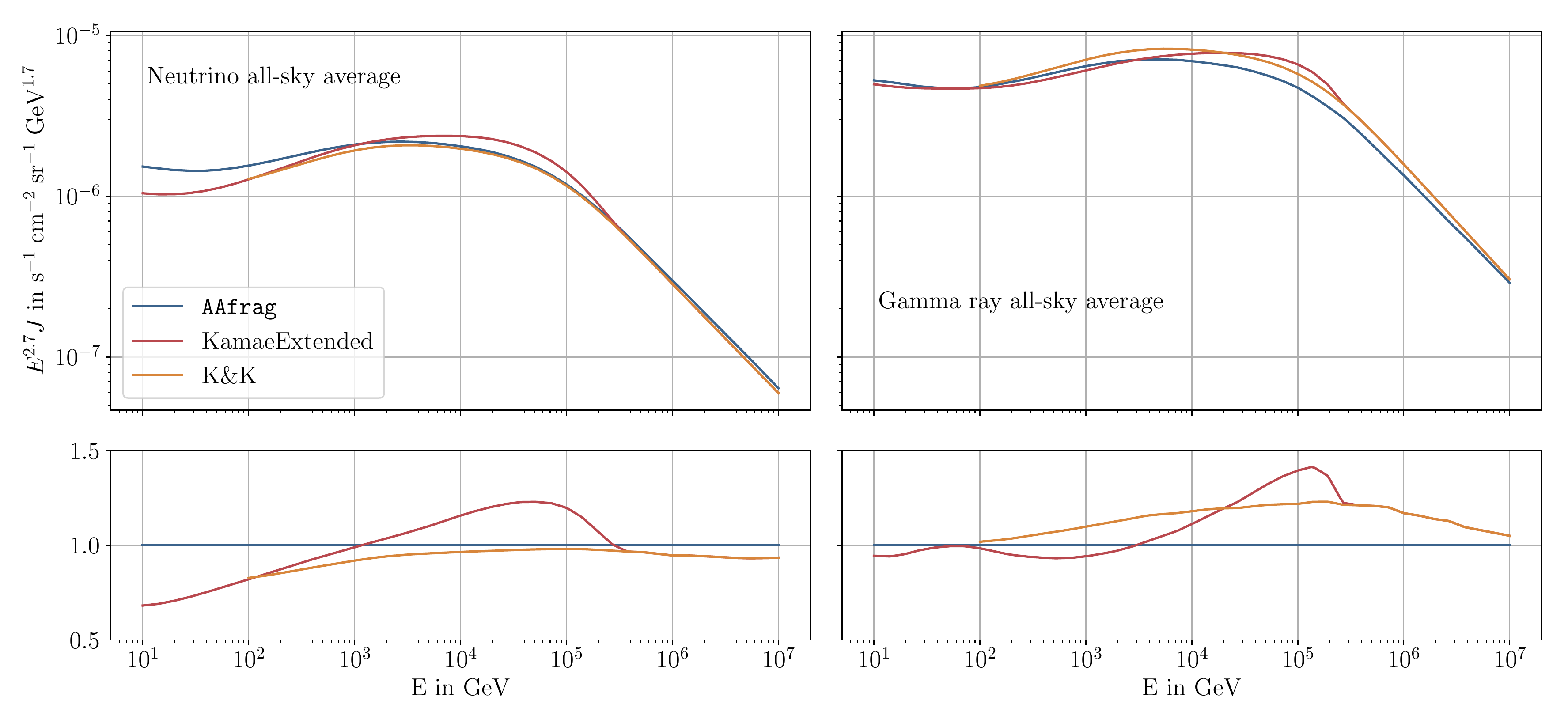}
\caption{
All-sky averaged neutrino (left panels) and hadronic gamma-ray (right panel) intensity as a function of energy. 
We show our model prediction for the different hadronic production cross-sections described in Sec.~\ref{sec:cross-sections} using the the best-fit GCR parameters combined with the \texttt{GALPROP} galactic gas maps and the \citet{Ferriere2001} source distribution. The blue line obtained using the \texttt{AAfrag} cross-sections corresponds to our fiducial model.
In the lower panels, the relative deviations after normalization to this fiducial model are presented.
}
\label{fig:xsec_profiles}
\end{figure*}

Figure~\ref{fig:xsec_profiles} shows the all-sky averaged spectra and relative differences of both the diffuse neutrino and hadronic gamma-ray flux for the three hadronic production cross-sections described in Sec.~\ref{sec:gas_maps}. All other inputs are the same as for our fiducial model. We remind the reader that the K\&K cross-sections are only valid at primary energies above $100\,\mathrm{GeV}$ and that the KamaeExtended model is equal to the K\&K model at primary energies above $500\,\mathrm{TeV}$. The latter is then reflected at somewhat lower energies in the diffuse emission spectra as is clearly visible in Figure~\ref{fig:xsec_profiles}. When comparing the differences between the cross-sections for neutrinos to those for hadronic gamma-rays, one can see that for hadronic gamma-rays, the cross-sections are in close agreement below $1\,\mathrm{TeV}$. This is different for neutrinos, where the difference between the \texttt{AAfrag} and KamaeExtended parametrizations are already sizable at lower energies. We also highlight that the KamaeExtended model predicts consistently harder spectra than the \texttt{AAfrag} model for both neutrinos and hadronic gamma-rays.

\subsection{Gas distribution}
\begin{figure*}[h]
\centering
\includegraphics{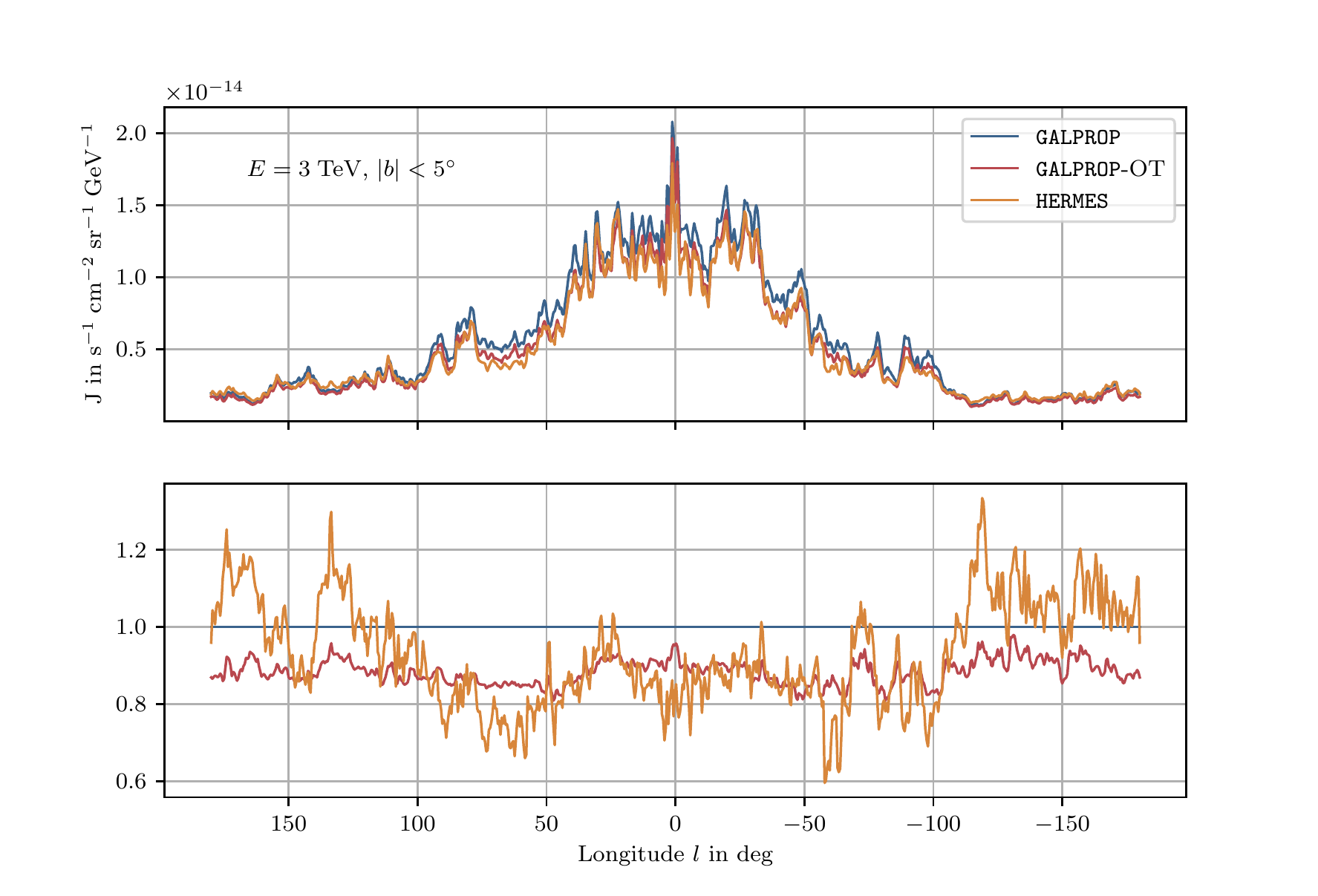}
\caption{
Neutrino intensity as a function of galactic longitude for a neutrino energy of $3 \, \text{TeV}$. 
We show our model prediction for the different gas maps described in Sec.~\ref{sec:gas_maps} using the the best-fit GCR parameters combined with the \citet{Ferriere2001} source distribution and the \texttt{AAfrag} production cross-sections. The blue profile obtained using the \texttt{GALPROP} galactic gas maps corresponds to our fiducial model.
In the lower panel, the relative deviations after normalization to this fiducial model are presented.
}
\label{fig:gas_map_profiles}
\end{figure*}

Figure~\ref{fig:gas_map_profiles} shows the longitudinal profiles of the diffuse neutrino flux at $E=3\;\mathrm{TeV}$ for latitudes $|b|<5^{\circ}$ for the three galactic gas maps described in Sec.~\ref{sec:gas_maps}. All other inputs are the same as for our fiducial model. The difference between the \texttt{GALPROP} and \texttt{GALPROP}-OT model arises solely from the assuption of a different value for the spin temeperature \Ts. As expected, the \texttt{GALPROP}-OT model assuming optically thin gas leads to a lower intenity. The differences with respect to the \texttt{HERMES} gas maps stem from the reliance on different $21\,\mathrm{cm}$ surveys, the reconstruction into galactocentric rings of different sizes, the assumed values of \Ts{} and the treatment of dark gas.  

\subsection{Cosmic ray source distribution}
\begin{figure*}[h]
\centering
\includegraphics{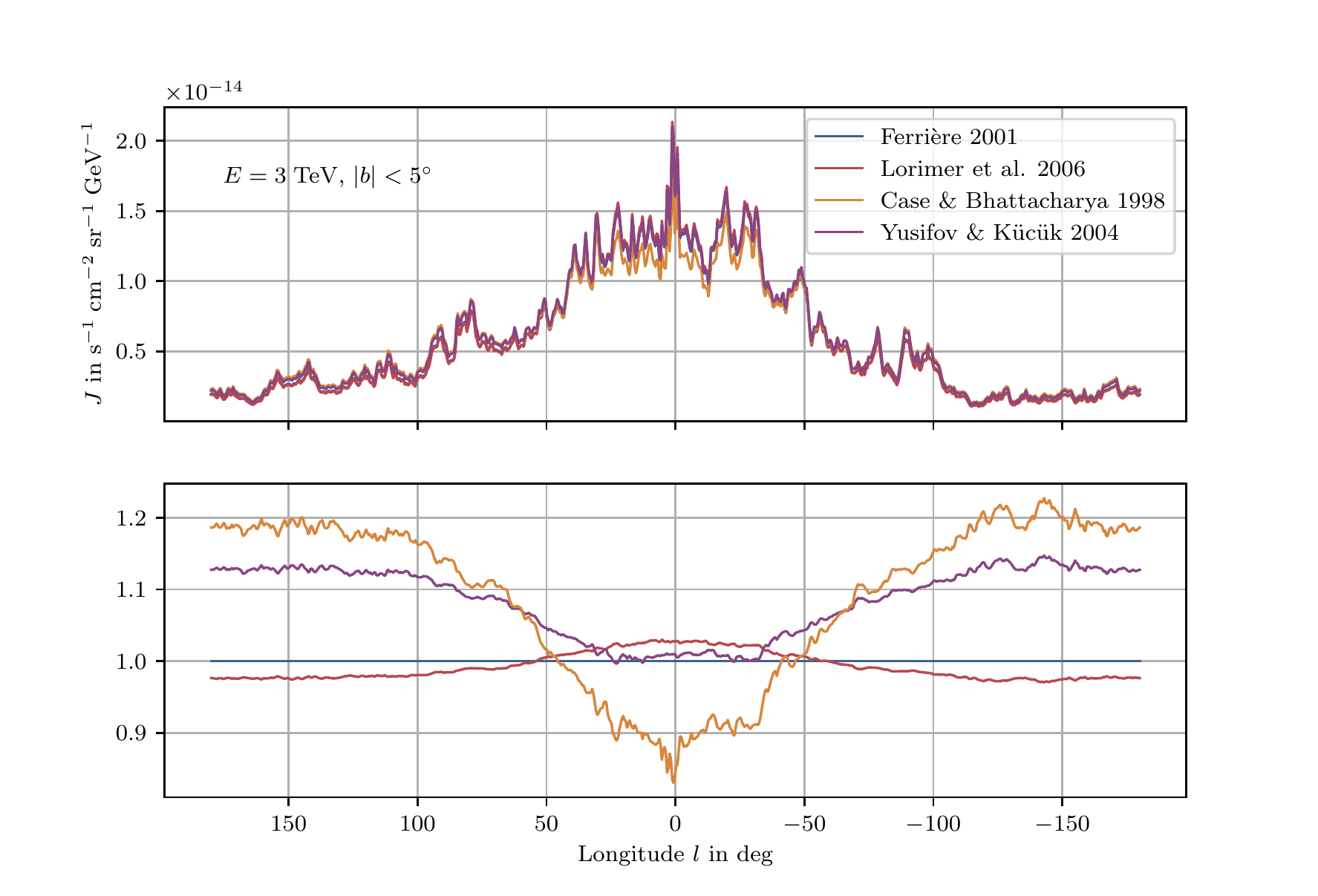}
\caption{
Neutrino intensity as a function of galactic longitude for a neutrino energy of $3 \, \text{TeV}$. 
We show our model prediction for the different source distributions described in Sec.~\ref{sec:GCR_model} and shown in Figure~\ref{fig:sd_radial_profiles} using the the best-fit GCR parameters combined with the \texttt{GALPROP} galactic gas maps and the \texttt{AAfrag} production cross-sections. The blue profile obtained using the \citet{Ferriere2001} source distribution corresponds to our fiducial model.
In the lower panel, the relative deviations after normalization to this fiducial model are presented.
}
\label{fig:sd_lon_profiles}
\end{figure*}

Figure~\ref{fig:sd_lon_profiles} shows the longitudinal profiles of the diffuse neutrino flux at $E=3\;\mathrm{TeV}$ for latitudes $|b|<5^{\circ}$ for the four source distributions described in Sec.~\ref{sec:source_injection}. All other inputs are the same as for our fiducial model. The relative intensities of the models towards and away from the galactic center reflect the shape of the radial profiles as shown in Figure~\ref{fig:sd_radial_profiles}. For the reasons described in Sec.~\ref{sec:diffuse_neutrino_intensities}, the relative differences between the intensities are larger away from the galactic center than towards it. Furthermore, as also mentioned in Sec.~\ref{sec:diffuse_neutrino_intensities}, the coincidence of the intensities at $|l|\approx 45^{\circ}$ can be explained by the fact that in this direction, the emission is on average produced at galacto-centric radii similar to the solar radius and is thus constrained by the fit to local GCR intensities. Overall, the source distribution from \citet{Case1998} gives the flattest emission profile, the emission obtained with the distribution from \citet{Lorimer2006} is peaked towards the galactic center the most.
\subsection{Interstellar radiation field}
\begin{figure*}[h]
\centering
\includegraphics{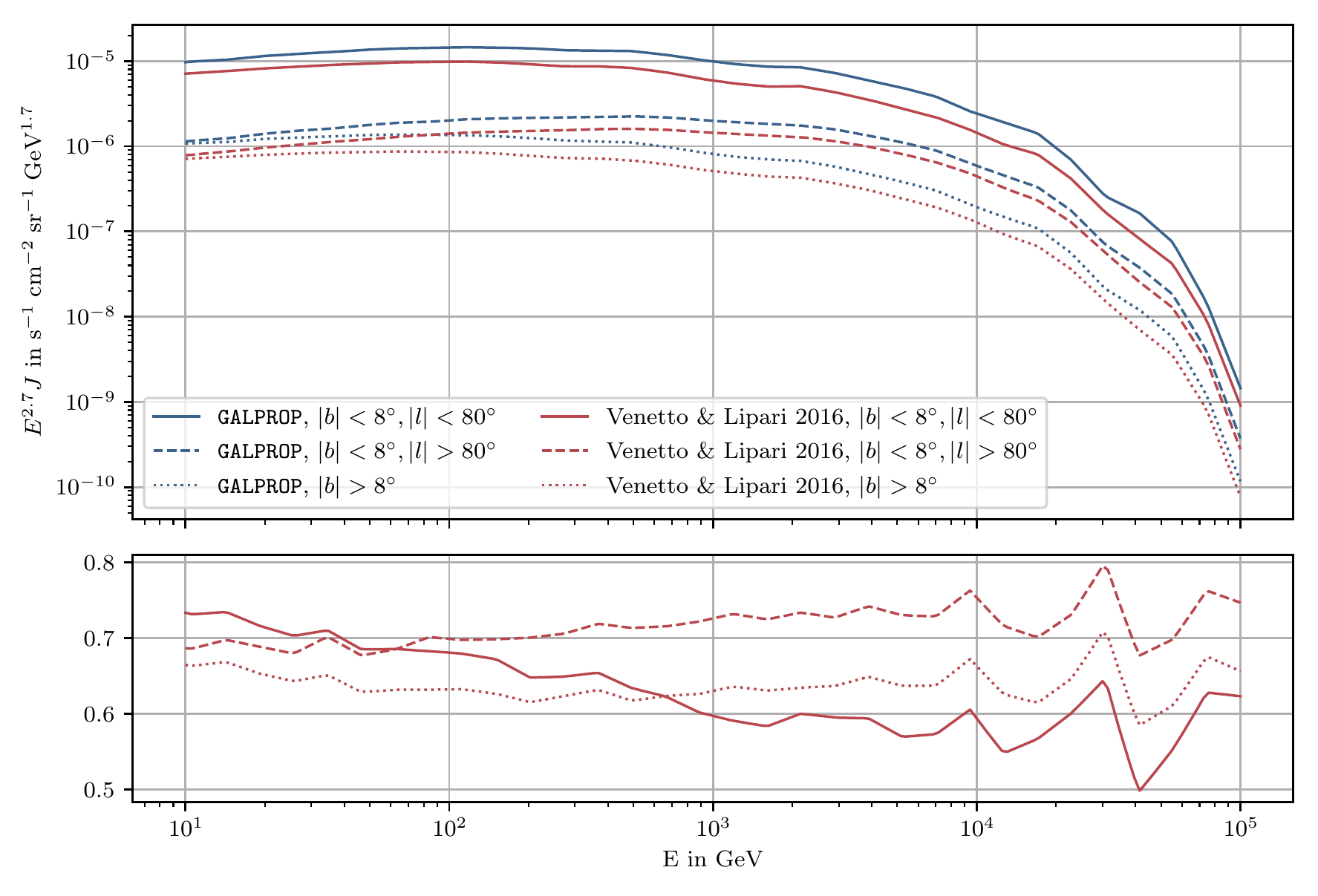}
\caption{
Inverse Compton gamma-ray intensity as a function of energy in different regions in the sky. 
We show our model prediction for the ISRF models described in Sec.~\ref{sec:ISRFs} using the the best-fit GCR parameters combined with the \citet{Ferriere2001} source distribution. The blue profiles obtained using the \texttt{GALPROP} ISRF model correspond to our fiducial model.
In the lower panels, the relative deviations after normalization to this fiducial model are presented.
}
\label{fig:ic_profiles}
\end{figure*}

Figure~\ref{fig:ic_profiles} shows the spectra of the Inverse Compton gamma-ray intensity in different windows in the sky. The spectra are calculated using the the best-fit GCR parameters combined with the \citet{Ferriere2001} and are shown for both ISRF models described in Sec.~\ref{sec:ISRFs}. Similar to the hadronic component of the gamma-ray and neutrino intensities, the Inverse Compton intensity also varies over the different windows in the sky, with the largest intensity towards the galactic center. The cut-off of the intensities above $10\,\mathrm{TeV}$ reflects the cut-off in the electron source spectra at $E^{e^{-}}_{\text{cut}}=20 \, \mathrm{TeV}$ as described in Sec.~\ref{sec:source_injection}. The \texttt{GALPROP} ISRF model produces intensities at least $25\,\%$ larger than the model from \citet{Vernetto2016} in all regions in the sky and at all energies. However, because the Inverse Compton intensity is a subdominant contribution to the overall gamma-ray intensity, the overall uncertainty stemming from the ISRF model choice as shown in Figures~\ref{fig:gamma_fiducial_gamma} and \ref{fig:GeV_gammas} is correspondingly reduced.
\end{appendix}
\end{document}